%% file: Manuscript_PMB-112620.tex
\documentclass[12pt]{iopart}

\usepackage{amsfonts}
\usepackage{amssymb}
\usepackage{amsopn} 
\usepackage{bm}

\usepackage[labelformat=simple]{subcaption}

\def\*#1{\bm{#1}}
\def\-#1{\mathcal{#1}}
\def\+#1{\mathrm{#1}}
\def\R{\mathbb{R}}

\newcommand{\transp}{^\top}

\newcommand{\ud}{\mathrm{d}}

\newcommand{\argmax}{\operatornamewithlimits{arg\,max}}
\newcommand{\argmin}{\operatornamewithlimits{arg\,min}}

\usepackage{cite}
\usepackage[ruled]{algorithm2e}
\usepackage{graphicx}
\usepackage{textcomp}
\usepackage{microtype} 
\usepackage{harvard}
\usepackage{acro}

\usepackage{tikz} 
\usepackage{pgfplots}
\pgfplotsset{compat = 1.16}

\usepackage{tikz}
\usetikzlibrary{spy}

\usepackage{chngcntr}
\counterwithout{equation}{section}

\tikzstyle{only in spy node magn 1.75}=[transform canvas={
	shift={(tikzspyinnode)}, scale=1.75}]

\usepackage{adjustbox}

\usepackage{pgf}
\usetikzlibrary{matrix}
\usetikzlibrary{decorations.pathmorphing} 
\usetikzlibrary{fit}					
\usetikzlibrary{backgrounds}
\usetikzlibrary{shapes.arrows}
\usetikzlibrary{shapes.geometric}
\usetikzlibrary{shapes.multipart}
\usetikzlibrary{matrix,arrows,decorations.pathreplacing,positioning}
\tikzstyle{rec}=[draw,rectangle, minimum height=2cm]
\tikzset{>=stealth', punkt/.style={rectangle, fill=gray!40, 
		draw=black, very thick, node distance=45pt, text width=5.5em, minimum height=3em, text centered}}
\tikzset{>=stealth', punkt_l/.style={rectangle, fill=gray!40, 
		draw=black, very thick, text width=3em, minimum height=2.5em, text centered}}
\tikzstyle{background}=[rectangle,fill=green!20,inner sep=0.1cm, rounded corners=4mm,minimum height=3em]
\tikzstyle{sum} = [draw, fill=gray!40, circle, node distance=1cm]
\tikzstyle{dot} = [circle, fill=black, inner sep=0pt, minimum size=5pt, node contents={}]
\tikzstyle{text_n} = [node distance=30pt, text width=8em, minimum height=2.5em, text centered]
\tikzstyle{text_upn} = [node distance=0pt, text width=12em, minimum height=2.5em, text centered]
\tikzstyle{fig_n} = [node distance=0pt, inner sep=0cm]

\include{acronyms}

\usepackage{etoolbox} 
\robustify\cite
\robustify\citeasnoun
\robustify\ref

\begin{document}

\title[]{Multi-Channel Convolutional Analysis Operator Learning for Dual-Energy CT Reconstruction}

\author{Alessandro~Perelli$^{\ast1, 2}$, Suxer~Alfonso~Garcia$^{\ast1}$, Alexandre~Bousse$^1$, Jean-Pierre Tasu$^3$, Nikolaos Efthimiadis$^3$, Dimitris~Visvikis$^1$}

\address{$^1$\ LaTIM, INSERM UMR 1101, Université de Bretagne Occidentale, 29238 Brest, France}
\address{$^2$\ School of Science and Engineering, University of Dundee, Scotland DD1 4HN, UK}
\address{$^3$\ Department of Radiology, University Hospital Poitiers, Poitiers, France} 
\address{$^\ast$\ These authors equally contributed to this work.} 
\ead{aperelli001@dundee.ac.uk}
\ead{suxer-lazara.alfonso-garcia@inserm.fr}
\vspace{10pt}

\begin{abstract}
	\emph{Objective.} \Ac{DECT} has the potential to improve contrast, reduce artifacts and the ability to perform material decomposition in advanced imaging applications. The increased number or measurements results with a higher radiation dose and it is therefore essential to reduce either number of projections per energy or the source X-ray intensity, but this makes tomographic reconstruction more ill-posed. 
	
	\emph{Approach.} We developed the \ac{MCAOL} method to exploit common spatial features within attenuation images at different energies and we propose an optimization method which jointly reconstructs the attenuation images at low and high energies with a mixed norm regularization on the sparse features obtained by pre-trained convolutional filters through the \ac{CAOL} algorithm. 
	
	\emph{Main results.} Extensive experiments with simulated and real \ac{CT} data were performed to validate the effectiveness of the proposed methods and we reported increased reconstruction accuracy compared to CAOL and iterative methods with single and joint \ac{TV} regularization. 
	
	\emph{Significance.} Qualitative and quantitative results on sparse-views and low-dose \ac{DECT} demonstrate that the proposed \ac{MCAOL} method outperforms both \ac{CAOL} applied on each energy independently and several existing state-of-the-art \ac{MBIR} techniques, thus paving the way for  dose reduction.
\end{abstract}

\acresetall

\input{introduction}

\input{review_CAOL}

\input{method_MCAOL}

\input{DECT_MCAOL}

\input{results}

\input{conclusion}

\ack
This work was funded by the French National Research Agency (ANR) under grant \texttt{ANR-20-CE45-0020}.

\makeatletter
\renewcommand{\@biblabel}[1]{}
\makeatother
\section*{References}

\bibliographystyle{dcu}
\bibliography{biblio}

\end{document}

%% file: acronyms.tex
\DeclareAcronym{CT}{
	short=CT, 
	long=computed tomography,
}

\DeclareAcronym{ET}{
	short=ET, 
	long=emission tomography,
}

\DeclareAcronym{PET}{
	short=PET, 
	long=positron emission tomography,
}

\DeclareAcronym{MRI}{
	short=MRI, 
	long=magnetic resonance imaging,
}

\DeclareAcronym{CAOL}{
	short=CAOL, 
	long=convolutional analysis operator learning,
}

\DeclareAcronym{MCAOL}{
	short=MCAOL, 
	long=multi-channel convolutional analysis operator learning,
}

\DeclareAcronym{MBIR}{
	short=MBIR, 
	long=model-based iterative reconstruction,
}

\DeclareAcronym{DECT}{
	short=DECT, 
	long= dual-energy computed tomography,
}

\DeclareAcronym{KVp}{
	short=KVp, 
	long= peak kilo-voltage,
}

\DeclareAcronym{CS}{
	short=CS, 
	long= compressed sensing,
}

\DeclareAcronym{TV}{
	short=TV, 
	long= total-variation,
}

\DeclareAcronym{PICCS}{
	short=PICCS,
	long= prior image constrained compressed sensing,
}

\DeclareAcronym{DL}{
	short=DL,
	long=dictionary learning,
}

\DeclareAcronym{CDL}{
	short=CDL,
	long= convolutional dictionary learning,
}

\DeclareAcronym{GT}{
	short=GT,
	long=ground truth,
}

\DeclareAcronym{XCAT}{
	short=XCAT,
	long=extended cardiac-torso,
}

\DeclareAcronym{MSE}{
	short=MSE,
	long=mean squared error,
}

\DeclareAcronym{KeV}{
	short=KeV,
	long= kilo-electron volt,
}

\DeclareAcronym{ROI}{
	short=ROI,
	long= region of interest,
}

\DeclareAcronym{MLTR}{
	short=MLTR,
	long= maximum-likelihood reconstruction for transmission,
}

\DeclareAcronym{RMSE}{
	short=RMSE,
	long=root mean squared error,
}

\DeclareAcronym{STD}{
	short=STD,
	long=standard deviation,
}

\DeclareAcronym{SNR}{
	short=SNR,
	long=signal-to-noise ratio,
}

\DeclareAcronym{1D}{
	short=1-D,
	long=1-dimensional,
}

\DeclareAcronym{2D}{
	short=2-D,
	long=2-dimensional,
}

\DeclareAcronym{3D}{
	short=3-D,
	long=3-dimensional,
}

\DeclareAcronym{EM}{
	short=EM,
	long=expectation-maximization,
}

\DeclareAcronym{ML}{
	short=ML,
	long=maximum-likelihood,
}

\DeclareAcronym{SPS}{
	short=SPS,
	long=separable paraboloidal surrogate,
}

\DeclareAcronym{PWLS}{
	short=PWLS,
	long=penalized weighted least-squares,
}

\DeclareAcronym{NLL}{
	short=NLL,
	long= negative log-likelihood,
}

\DeclareAcronym{LBFGS}{
	short=L-BFGS,
	long=limited-memory Broyden-Fletcher-Goldfarb-Shanno,
}

\DeclareAcronym{BPEGM}{
	short=BPEG-M,
	long=\emph{Block Proximal Extrapolated Gradient method using a Majorizer},
}

\DeclareAcronym{PMOC}{
	short=PMOC,
	long=\emph{proximal mapping with orthogonality constraints},
}

\DeclareAcronym{JTV}{
	short=JTV,
	long=joint total variation,
}

\DeclareAcronym{CONVOLT}{
	short=CONVOLT,
	long=CONVolutional Operator Learning Toolbox,
}

\DeclareAcronym{ART}{
	short=ART,
	long=algebraic reconstruction technique,
}

\DeclareAcronym{POCS}{
	short=POCS,
	long=projection onto convex sets,
}

\DeclareAcronym{FWHM}{
	short=FWHM,
	long=full width at half maximum,
}

%% file: introduction.tex
\section{Introduction}\label{sec:introduction}

\Acf{DECT} is an energy-resolved X-ray imaging technique that offers two sets of attenuation  measurements  utilizing  two  different  energy spectra acquired over the same anatomical region \cite{zhang2020multi}. Since its invention, the technique has progressively been used for material decomposition and energy-selective imaging \cite{dong2014combined}. 
Applications of \ac{DECT} include differentiation and quantification of materials, tissue characterization, virtual monoenergetic imaging, automated bone removal in \ac{CT} angiography, cardiovascular imaging, multiple contrast agent imaging and mapping of effective atomic number, and radiotherapy  \cite{mccollough2015dual,forghani2018characterization,yu2012mono,van2016dual}. 

\Ac{DECT} acquisition techniques require two helical scans at two different tube voltages; therefore, two sets of projection data at different energy levels are collected and further reconstructed. 
However, as the number of incident photons increases when irradiating with two sources the same anatomical region, the radiation dose increases proportionally \cite{sajja2020technical}. 

A reduction in radiation exposure as well as in acquisition time can be achieved by decreasing the number of projection angles and/or reducing the X-ray source dose. 
However, aliasing artifacts can appear in the reconstructed images if the number of projection angles does not follow the Nyquist sampling theorem. Moreover, it is more challenging to achieve high-resolution, high-contrast image reconstruction due to the low \ac{SNR} \cite{zhang2020multi}. 

In the literature, most of the development on low-dose \ac{CT} reconstruction has focused on single image. Among the main techniques, \ac{MBIR} methods are the most popular. These techniques exploit models of the imaging system's physics (forward models) along with statistical models of the measurements and noise and often simple object priors. They iteratively optimize model-based cost functions to estimate the underlying unknown image \cite{Elbakri2002}. Typically, such cost functions consist of a data-fidelity term, e.g., least squares or \ac{NLL}, capturing the imaging forward model and the measurement/noise statistical model and a regularizer term promoting smoothness, low-rank or sparsity \cite{kim2014sparse}. 
\Ac{TV} \cite{Sidky2006Accurate,sidky2008image} has been proposed to solve incomplete projection data reconstruction problems and achieved good performance. However, \ac{TV} assumes that the signal is piecewise constant, resulting in undesired patchy effects on the reconstructed images \cite{block2007undersampled}. 
Recent developments and interests within spectral \ac{CT} refer to self-supervised methods using only noisy data \cite{fang2021} and dynamic \ac{DECT}, which refers to randomly changing the energy threshold of the detectors to obtain the spectral information \cite{yao2019}.

Data-driven and learning-based approaches have gained much interest in recent years for biomedical image reconstruction. These methods learn representations of images and are used in combination with \ac{MBIR} techniques to perform complex mappings between limited or corrupted measurements and high-quality images. Among those algorithms, data-driven sparse transforms such as \ac{DL} \cite{xu2012} use a training dataset of high-resolution and denoised images to learn features, in an unsupervised manner, that can be used to reconstruct new images. These features take the form of ``atoms'', which are regrouped into dictionaries and are used to sparsely represent the image \cite{aharon2006k}. \Ac{DL}-based image reconstruction integrates the learned atoms with the raw scanner data within a regularized \ac{MBIR} context \cite{ravishankar2017efficient,zheng2018pwls}. Other closely related methods include sparsifying transform learning \cite{ravishankar2012} and the connection between data-adaptive models and convolutional deep learning algorithms \cite{ravishankar2019} with an increase of interest in methods that leverage both learning-based and \ac{MBIR} tools.

However, most \ac{DL} methods are patch-based, and the learned features often contain shifted versions of the same features. The resulting learned dictionary may be redundant and therefore are memory demanding, which makes it difficult to utilize in \ac{3D} multimodal imaging. 
To address these problems, \ac{CDL} techniques utilize shift-invariant filters, providing a convenient and memory-efficient alternative to conventional \ac{DL} techniques \cite{Chun2017}. \Ac{CDL} approaches can be combined with \ac{MBIR} by providing unsupervised prior knowledge of the target image. 
The \ac{CDL} approach can also be formulated from an analysis point of view \cite{Chun2019} (sparse convolution) and is known as \ac{CAOL}.

Despite the rapidly expanding research, the application of \ac{CDL} to multi-channel images has received little attention \cite{degraux2017online,garcia2018convolutional_p}. 
However, image reconstruction from \ac{DECT} sparse-views or low-dose requires algorithms more advanced than the standard approach where attenuation at each measured energy is reconstructed independently. Notable models in the literature designed to promote structural similarity of images are \ac{JTV} \cite{ehrhardt2014,cueva2021synergistic}, spectral patch-based penalty for the maximum-likelihood method \cite{kim2015patch}, tensor-based and coupled dictionary learning \cite{wu2018dictionary,song2019coupled}, parallel level sets \cite{kazantsev2018joint} and the prior rank, intensity and sparsity model (PRISM) \cite{yang2017prism}.

\subsection{Main Contribution} 

We extend the CAOL approach to multichannel settings and we develop a \ac{MCAOL} framework that can exploit direct joint reconstruction, given the low-dose \ac{DECT} measurements,  where all the unknown images are reconstructed simultaneously by solving one combined optimization problem. This is the first time that MCAOL is applied to DECT image reconstruction and we demonstrate its superiority respect to CAOL. Furthermore, MCAOL requires considerably less memory compared to alternative DL approaches. 
The joint reconstruction approach is developed for a low-dose data acquisition protocol which consists of collecting data using a sparse angular sampling, using a different X-ray energy in consecutive steps and low X-ray photon counts. 

In \ac{DECT}, a reasonable prior assumption is that attenuation images at different energies can be expected to be {\em structurally} similar in the sense that an edge (e.g., an organ boundary) that is present at one energy, is likely to be at same location and alignment with the other energies as well, even though the contrast between materials will be different at each energy. 

MCAOL technique reconstructs attenuation images from the projection data combined with multi-channel filters trained on a dataset of reconstructed images. 

The central idea of \ac{MCAOL} is to learn unsupervised \ac{DECT} muti-channel convolutional dictionaries that can provide a joint sparse representation of the underlined images by jointly learning filters for the different energies: each atom not only carries individual information for each energy individually but also inter-energy information. 

By reconstructing \ac{DECT} images using \ac{MBIR} techniques in conjunction with \ac{MCAOL}, the multi-energy information can be optimally used by allowing the images to ``talk to each other'' during the reconstruction process through the learned joint dictionaries, reducing noise while preserving image resolution. In order to deal with the extreme low-dose scenario, we model the Poisson model exactly and we solve the image optimization problem by using approximated quasi-Newton method with constrained memory to achieve accurate joint reconstruction with limited computational complexity.

\subsection{Notation and Paper Organization}

The remainder of this paper is structured as follows: Section~\ref{sec:CAOL} briefly reviews the single-channel \ac{CAOL} algorithm as initially proposed in \cite{Chun2019} while Section~\ref{sec:MCAOL} presents the proposed approach to extend \ac{CAOL} to multi-channel problems.  In Section~\ref{sec: DECT_recon} we first describe the physical model of X-ray \ac{DECT} from the continuous to discrete domain and the Poisson noise model; furthermore, we illustrate the proposed method to minimize the \ac{MCAOL} reconstruction objective function using the exact low-dose \ac{DECT} model. Finally, in Section~\ref{sec:results} comprehensive results of \ac{MCAOL} on a numerical \ac{XCAT} phantom \cite{segars2008xcat} and clinical acquisition of a patient's full body are shown together with a comparison of its performance with state-of-the-art algorithms for both the sparse-views and low-dose \ac{DECT} reconstruction.

Throughout the paper we adopt the following notations: matrices or discrete operators and column vectors are written respectively in capital and normal boldface type, i.e. $\*A$ and $\*a$, to distinguish from scalars and continuous variables written in normal weight; $[\*a]_j$ denotes the $j$-th entry of $\*a$; `$(\cdot)\transp$' denotes the transposition; `$\circledast$' denotes the \ac{2D} convolution operator; given an image  $\*x\in\R^J$ and a filter ${\*d} \in\R^P$, $P<J$, $\*d \circledast \*x$ represents the \ac{2D} convolution obtained by zero padding and then truncating the boundary such that the output dimension is the same as the input dimension $J$; an image $\*x\in\R^J$ is represented by a column vector for algebraic operations and by a $\sqrt{J} \times \sqrt{J}$ matrix for \ac{2D} convolutions. Finally, we indicate the set of positive real numbers with the notation $]0, +\infty[$.

%% file: review_CAOL.tex
\section{Learning Convolutional Regularizers for Image Reconstruction: CAOL}\label{sec:CAOL}

In this Section we review the foundation of \ac{CAOL} for \ac{MBIR} which is achieved by solving an optimization problem of the form
\begin{equation}\label{eq:PML}
	\min_{\*x \in \R^J} L(\*x,\*y) + \beta R(\*x)
\end{equation} 
where $\*x \in \R^J$ is the \ac{2D} 
image to reconstruct, $\*y\in\R^I$ is the observed measurement, $L$ is a data-fidelity term that incorporates the measurement model---generally taking the form of a \ac{NLL} function---and $R$ is a regularizer weighted by $\beta>0$; $I$ and $J$ are respectfully the dimension of the measurement (number of detectors) and dimension of the image (number of pixels). The minimization is carried out with the help of iterative algorithms such as modified \ac{EM} for \ac{ET} \cite{DePierro1995} or \ac{PWLS} combined with \ac{SPS} for \ac{CT} \cite{Elbakri2002}.

The regularizer $R$ is designed such that the reconstructed image $\hat{\*x}(\*y)$ has desired properties, such as smoothness and sparsity of the gradient. It can be also trained so that $\hat{\*x}(\*y)$ can be sparsely represented as a linear combination of basic elements, or atoms, regrouped in a dictionary.

We consider the \ac{CAOL} approach \cite{Chun2019} where the image is sparsely represented with convolutional kernels (filters). In the analysis model, the image is represented with ``sparsifying'' filters ${\*d}_k \in\R^P$ by the analysis operator $\-A_{\*D}: \*x \mapsto \{\*d_k \circledast \*x\}$, such that
\begin{equation}
	\*d_k \circledast \*x \approx \*z_k, \quad  \forall k = 1, \dots, K .
\end{equation}
where $\*z_k\in \R^J$ is a sparse feature image vector of the same dimension as the image $\*x$. The filters $\*d_k \in\R^P$ are vectorized images of dimension $P \ll J$ that are regrouped in a dictionary $\*D = \{\*d_k\}\in\R^{P\times K}$.  

Learning the dictionary $\*D$ from a dataset of training images $\{\*x_l\in\R^J: l = 1,\ldots ,L\}$ corresponds to finding a collection of filters $\*D^{\star} = \{\*d^{\star}_k\}$ obtained by the following non-convex optimization problem
\begin{equation}\label{eq:d}
	\*D^{\star} = \argmin_{\*D \in C} \, \min_{\{\*z_{l,k}\}} F_{\+a} (\*D, \{\*z_{l,k}\})
\end{equation}

\noindent where $C$ is the constraints on $\*D = \{\*d_k\}$ and with the training analysis objective function $F_\+a$ defined as
\begin{equation}\label{eq:cost_tr_CAOL}
	F_{\+a}(\*D, \{\*z_{l,k}\}) = \sum_{l=1}^{L} \sum_{k=1}^{K} \frac{1}{2} \left\| \*d_k \circledast \*x_l  - \*z_{l,k}\right\|_2^2 + \alpha\left\| \*z_{l,k} \right\|_0
\end{equation} 
where $\*z_{l,k}\in \R^J$ is the feature image associated to the training image $\*x_l$ and the filter $\*d_k$, $\left\|\cdot\right\|_0$ is the sparsity-promoting $l_0$ semi-norm defined for all $\*z=[z_1,\dots,z_J]\transp \in \R^J$ as
\begin{equation}
	\|\*z\|_0 = \sum_{j=1}^J \bm{1}_{]0,+\infty[}(|z_j|)
\end{equation}
where $\bm{1}_A\colon\mathbb{R}\rightarrow\{0,1\}$ denotes the indicator function of a set $A\subset \mathbb{R}$, which is defined as $\*1_A(\xi)=1$ if $\xi\in A$ and $\*1_A(\xi)=0$ if $\xi\notin A$, and $\alpha > 0$ is a weight balancing between accuracy and sparsity. In \citeasnoun{Chun2019} the filters are enforced to satisfy the \textit{tight-frame} conditions, i.e.,
\begin{equation}
	C = \left\{ \{\*d_k\}: [\*d_1, \ldots , \*d_K] [\*d_1, \ldots , \*d_K]\transp = \frac{1}{P}\*I_K\right\} 
\end{equation}
where $\*I_K$ is the $K\times K$ identity matrix, to promote filters diversity. The optimization problem~\eref{eq:d} is solved by alternating between minimization in $\*z$  and in $\*D$. The joint minimization in $\*D$, $\*z$ can be achieved with a \ac{BPEGM} algorithm, which can be implemented using the \ac{CONVOLT} \cite{Chun2019,convolt}. BPEG-M alternates between minimization in $\*z$ and $\*D$. The minimization in $\*z$ is obtained with a hard-thresholding operator $\-T\colon\R^J\times\R_+^*\to\R^J$ defined at each row $j$ as
\begin{equation}\label{eq:thres}
	[\-T(\*a,\beta)]_j = \cases{a_j  &for $\frac{1}{2} a_j^2  \ge \beta $\\
		0 &otherwise\\}
\end{equation}
for all $\*a =  [a_1,\dots,a_J]\transp \in \R^J$ and for all $\beta >0$, which provides a global minimizer for $\*z\mapsto \frac{1}{2} \| \*a - \*z  \|_2^2  +   \beta \|\*z\|_0$, in such a way that
\begin{equation}
	\mathcal{T}(\*d_k \circledast \*x_l , \alpha) = \argmin_{\*z_{l,k}} \frac{1}{2} \| \*d_k \circledast \*x_l  - \*z_{l,k} \|_2^2 + \alpha\| \*z_{l,k} \|_0 \, .
\end{equation}

The minimization in $\*D$ is achieved with a \ac{PMOC}, which can be implemented with a singular value decomposition \cite{Chun2019}. Finally the regularizer $R$ in the minimization problem~\eref{eq:PML} is derived from the learned filters $\*D^\star$ as 
\begin{equation}\label{eq:pen_CAOL}
	R(\*x) = \min_{\{\*z_k\}}\, \sum_{k=1}^{K} \frac{1}{2} \left\| \*d_k^\star \circledast \*x  - \*z_k\right\|_2^2 + \alpha\left\| \*z_k \right\|_0  \, .
\end{equation}

%% file: method_MCAOL.tex
\begin{figure}[!t]
	
	\vspace{-.5cm}
	\begin{adjustbox}{max totalsize={\textwidth}{\textheight},center}
		\begin{tikzpicture}
			
			\node[text_n, inner sep=0pt, node distance=0cm,  text width=12em] at (0,0) (tr_text) {Training DECT dataset \newline $\*x_{e,l} : e = 1, 2 ; l = 1, \ldots , L$};
			
			\node[fig_n, above= 0.1 of  tr_text, inner sep=0pt] (tr_im1) 
			{\includegraphics[width=.25\textwidth]{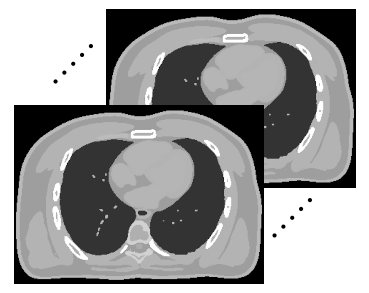}};
			
			\node[fig_n, below= 0.1 of  tr_text, inner sep=0pt] (tr_im2) 
			{\includegraphics[width=.25\textwidth]{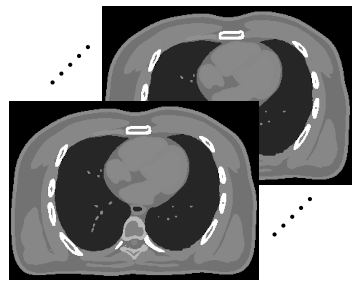}};
			
			\node [punkt,right= of  tr_text, label=above:MCAOL from $\{\*x_{e,l}\}$] (Tr_module)  {};
			
			\path[->] (tr_text.east) edge node {} (Tr_module.west);
			
			\node[text_n, right= 0cm and 1.5cm of  Tr_module] (filt_text) {Trained filters\\ $\*D^* = \{\*D^*_1, \*D^*_2\}$};
			
			\path[->] (Tr_module.east) edge node {} (filt_text.west);
			
			\node[fig_n, below right= 0cm and -.4cm of  filt_text, inner sep=0pt] (acq_CT) 	{\includegraphics[width=.25\textwidth]{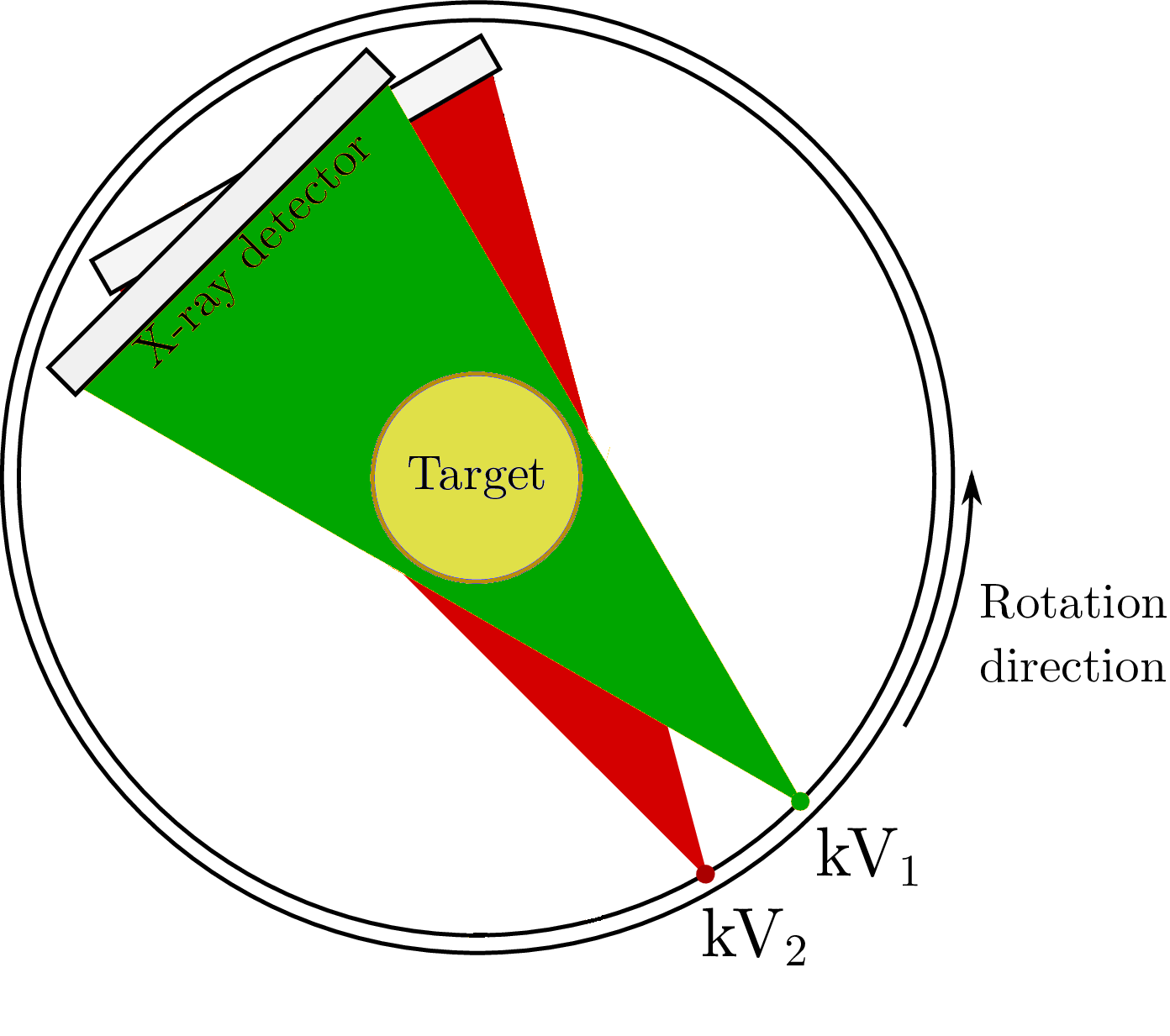}};
			
			\node[text_upn, below= of acq_CT, inner sep=0pt, node distance=0cm] (acq text) {\ac{DECT} acquisition};
			
			\node[fig_n, above= 0.1cm and 0cmof  filt_text, inner sep=0pt] (filt_im) 	{\includegraphics[width=.2\textwidth]{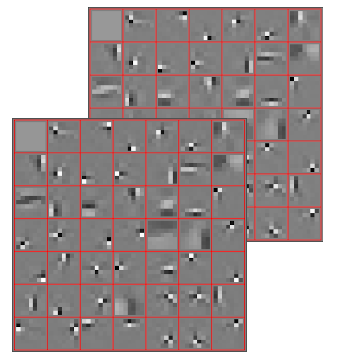}};
			
			\node [punkt,right= 0cm and 7cm of filt_text] (Rec_module)  {};	
			\path[->] (filt_text.east) edge node {} (Rec_module.west);
			
			\node[text_upn, above= of  Rec_module] (recn_text) {Reconstruction: apply $\*D^*$\\to MBIR DECT};
			
			\node[fig_n, below right= 0.1cm and -4.5cm of  Rec_module, inner sep=0pt] (sino_im) 	{\includegraphics[width=.25\textwidth]{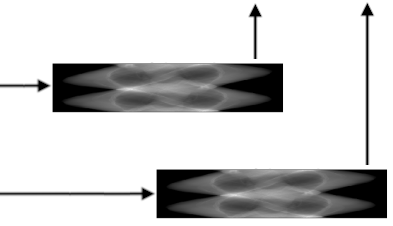}};
			
			\node[text_upn, below= of  sino_im, inner sep=0pt, node distance=0cm] (sino text) {Sinograms $\*y_e, e=1,2$};
			
			\node[text_n, right= of  Rec_module] (rec_text) {Reconstructed images\\ $\hat{\*x}_e(\*y_e), e = 1,2$};
			\path[->] (Rec_module.east) edge node {} (rec_text.west);
			
		\end{tikzpicture}
	\end{adjustbox}
	\vspace{-.7cm}
	\caption{Diagram of \ac{MCAOL} consisting of the unsupervised filter learning phase and the model-based iterative \ac{DECT} reconstruction module.}
	\label{fig:MCAOLfig}
\end{figure}
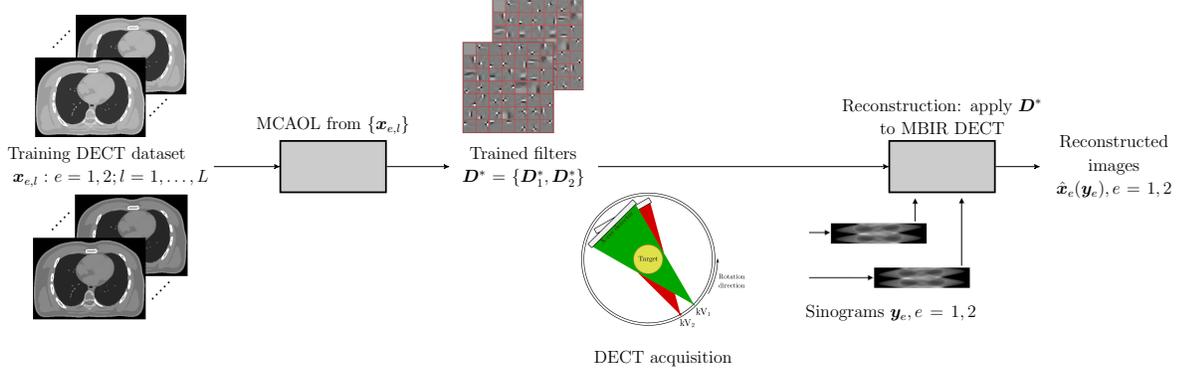

\section{Multi-channel Convolutional Analysis Operator Learning}\label{sec:MCAOL}

\Ac{MBIR} can be generalized to multi-channel imaging. Assuming we wish to reconstruct two images  $\*x_1,\*x_2\in\R^J$ of the same ``object''  from two independent measurements $\*y_1\in\R^{I_1}$ and $\*y_2\in\R^{I_2}$ corresponding to two modalities, multi-channel \ac{MBIR} can be achieved by using an iterative algorithm to solve   
\begin{equation}\label{eq:MPML}
	\min_{\*x_1,\*x_2\in\R^J} \rho_1 L_1(\*x_1,\*y_1) + \rho_2 L_2(\*x_2,\*y_2) + R_\mathrm{mc}(\*x_1,\*x_2)
\end{equation}
where $L_1$ and $L_2$ are the data-fidelity terms for $\*x_1$ and $\*x_2$, $R_\mathrm{mc}$ is a multi-channel regularizer and $\rho_1,\rho_2>0$ are weights. $R_\mathrm{mc}$ is designed to exploit the inference between the 2 channels $\*x_1$ and $\*x_2$, for example to promote structural similarities as proposed in \citeasnoun{ehrhardt2014}. 

\Ac{MCAOL} is a generalization of \ac{CAOL} where the training is performed jointly on a set of images obtained from imaging modalities as depicted in Fig.~\ref{fig:MCAOLfig} for \ac{DECT}. Let $\{(\*x_{1,l}, \*x_{2,l}) \in\R^J \times\R^J : l= 1,\ldots ,L\}$ be a training dataset consisting of $L$ pairs of images. 
\Ac{MCAOL} learns the sparsifying filter pairs 
\begin{equation}\label{eq:dual_d_def}
	(\*d_{1,k}, \*d_{2,k}) \in\R^P \times\R^P : k= 1,\ldots ,K
\end{equation}
together with the extracted feature pairs
\begin{equation}\label{eq:dual_z}
	(\*z_{1,l,k}, \*z_{2,l,k}) \in\R^J \times\R^J : k =1,\ldots , K,\;  l = 1,\ldots ,L \, .
\end{equation}
\Ac{MCAOL} is achieved by solving the following optimization problem, given the training image set $(\*x_{1,l}, \*x_{2,l})$
\begin{equation}\label{eq:dual_d_opt}
	( \*D_1^{\star}, \*D_2^{\star} )  
	= \argmin_{\*D_1, \*D_2, \in C} \min_{\{\*z_{1,l,k}\} \atop \{\*z_{2,l,k}\}} F_{\+{mc}} \left(\*D_1, \*D_2, \{\*z_{1,l,k}\}, \{\*z_{2,l,k}\}\right) 
\end{equation}
with the training cost function $F_{\+{mc}}$ defined as
\begin{eqnarray}
	F_{\+{mc}} (\*D_1, \*D_2, \{\*z_{1,l,k}\}, \{\*z_{2,l,k}\})   = \sum_{l=1}^{L} \sum_{k=1}^{K} \frac{\gamma_1}{2} \left\| \*d_{1,k} \circledast \*x_{1,l}  - \*z_{1,l,k}\right\|_2^2 \nonumber  \\ 
	+ \frac{\gamma_2}{2} \left\| \*d_{2,k} \circledast \*x_{2,l}  - \*z_{2,l,k}\right\|^2 + \left\| (\*z_{1,l,k}, \*z_{2,l,k}) \right\|_{1,0} \label{eq:Fma}
\end{eqnarray}
where $\gamma_1, \gamma_2 > 0$ are weights and the semi-norm $\|\cdot \|_{1,0}$ on $\R^J \times \R^J$ is defined for all $\*z_1 = [z_{1,1},\dots,z_{1,J}]\transp\in\R^J$ and for all $\*z_2 = [z_{2,1},\dots,z_{2,J}]\transp\in\R^J$ as
\begin{equation}\label{eq:l_1_0}
	\| (\*z_1, \*z_2) \|_{1,0} = \sum_{j=1}^J\*1_{]0, +\infty[}\left(|z_{1,j}| + |z_{2,j}|\right) 
\end{equation}
$\|\cdot \|_{1,0}$ denotes the $l_{1,0}$ norm. It promotes joint sparsity, i.e., with zero and non-zero values at the same locations, of image features in all the modalities, that are encoded by the multi-channel dictionary $\*D_1, \*D_2$.  

To solve (\ref{eq:dual_d_opt}) we utilize an iterative alternating minimization scheme where the update of the filters $( \*D_1, \*D_2 )$ is solving a block non-convex problem using the \ac{PMOC} algorithm \cite{Chun2019,convolt} while for the update of the sparse codes $(\*z_{1,l,k}, \*z_{2,l,k})$ we deploy a multi-channel hard-thresholding operator  $\-T_{\+{mc}}\colon\R^J\times \R^J\times (\R_+^*)^2 \to \R^J\times \R^J$ defined at each row $j$ as 
\begin{equation}\label{eq:mthres}
	[\-T_{\+{mc}}(\*a_1,\*a_2,\*\gamma)]_j = \cases{(a_{1,j},a_{2,j})  & for $\frac{1}{2}\gamma_1 a_{1,j}^2 + \frac{1}{2}\gamma_2 a_{2,j}^2 \ge 1$\\
		(0,0) &otherwise\\}
\end{equation}
for all $\*a_1  =  [a_{1,1},\dots,a_{1,J}]\transp \in \R^J$, $\*a_2  =  [a_{2,1},\dots,a_{2,J}]\transp \in \R^J$  and  for all $\*\gamma = (\gamma_1,\gamma_2)\in(\R_+^*)^2$, which provides a global minimizer for $(\*z_1,\*z_2) \mapsto \frac{\gamma_1}{2} \| \*a_1 - \*z_1 \|_2^2  + \frac{\gamma_2}{2} \| \*a_2 - \*z_2 \|_2^2 + \|\*z_1,\*z_2\|_{1,0} $ \cite[Section 3]{xu2011image}, in such a way that
\begin{eqnarray}
	\-T_{\+{mc}}(\*d_{1,k} \circledast \*x_{1,l}, \*d_{2,k} \circledast \*x_{2,l}, \*\gamma  ) =  \argmin_{\*z_{1,l,k},\*z_{2,l,k}}  \, \Bigg\{ \frac{\gamma_1}{2} \| \*d_{1,k} \circledast \*x_{1,l}  - \*z_{1,l,k}\|_2^2 \nonumber  \\ 
	+ \frac{\gamma_2}{2} \| \*d_{2,k} \circledast \*x_{2,l}  - \*z_{2,l,k}\|^2 + \| (\*z_{1,l,k}, \*z_{2,l,k}) \|_{1,0}   \Bigg\}   \, .
\end{eqnarray}

Finally the regularizer $R_{\+{mc}}$ in the minimization problem~\eref{eq:MPML} is derived from the learned filters $(\*D_1^\star,\*D_2^\star)$ as 
\begin{eqnarray}
	R_{\+{mc}}(\*x_1,\*x_2)  =   \min_{\{\*z_{1,k}\} \atop \{\*z_{2,k}\}} \sum_{k=1}^{K} \frac{\gamma_1}{2} \left\| \*d^\star_{1,k} \circledast \*x_1  - \*z_{1,k}\right\|_2^2 \nonumber \\
	\quad + \frac{\gamma_2}{2} \left\| \*d^\star_{2,k} \circledast \*x_{2,l}  - \*z_{2,k}\right\|_2^2 + \left\| (\*z_{1,k}, \*z_{2,k}) \right\|_{1,0} \label{eq:pen_MCAOL}
\end{eqnarray}

\noindent The pseudo-code for \ac{MCAOL} training procedure is summarized in Algorithm~\ref{algo:MCAOL Training}.

\begin{algorithm}[!t]
	\SetAlgoLined
	\KwIn{DE Training Dataset $\*x_{e,l}$, $l=1,\ldots ,L$, $e = 1,2$, joint sparsity weights $\bm\gamma = (\gamma_1,\gamma_2)$, \#outer iterations $N_\mathrm{outer}$}
	\KwOut{Learned filters $( \*D_1^{\star}, \*D_2^{\star} )$}
	$(\*D_1^0, \*D_2^0) \leftarrow$ $\+{Normalized}$ $\+{random}$ $\+{initialization}$ \;
	
	\For{$t = 0, \ldots, N_\mathrm{outer}-1$}{
		\textit{Update sparse codes (in parallel)} \;
		\For{k,l=1,1,\dots,K,L}{
			$(\*z^{t+1}_{1,l,k}, \*z^{t+1}_{2,l,k}) \leftarrow$ $\mathcal{T}_{\+{mc}}(\*d^{t+1}_{1,k} \circledast \*x_{1,l}, \*d^{t+1}_{2,k} \circledast \*x_{2,l}, \bm\gamma)$ \;
		}
		\textit{Update Filters } \;
		$ \*D^{t+1}_{1} \leftarrow \mathrm{PMOC}(\*x_{1,l},\*z^{t+1}_{1,k})$ \;
		$ \*D^{t+1}_{2} \leftarrow \mathrm{PMOC}(\*x_{2,l},\*z^{t+1}_{2,k})$ \;
		
	}
	$\*D_1^{\star} \leftarrow  \*D_1^{N_\mathrm{outer}} $ \; 
	$\*D_2^{\star} \leftarrow  \*D_2^{N_\mathrm{outer}} $ \; 
	\caption{\ac{MCAOL} Training Algorithm}\label{algo:MCAOL Training}
\end{algorithm}

%% file: DECT_MCAOL.tex
\section{Dual-Energy CT Reconstruction with Multi-Channel \ac{CAOL}}\label{sec: DECT_recon}

\subsection{X-ray \ac{CT} Discrete Model}

In this section, we describe the \ac{CT} discrete physical measurement process with the spectrum of the X-ray source beams composed of two different energies. We consider the case of \ac{2D} slice-by-slice imaging systems.

For image reconstruction we assume that the continuous attenuation image $\mu_e(\*r)$ which denotes the linear attenuation coefficient at position $\*r\in \R^2$ and the energy level $e=1,2$, can be represented by a linear combination of basis functions $\{b_j\}$  associated to a discrete sampling on a $\sqrt{J} \times \sqrt{J}$ Cartesian grid,
\begin{equation}\label{eq:paramdiscrete}
	\mu_e(\*r) = \sum_{j=1}^{J} x_{e,j} b_j(\*r) \, ,
\end{equation}
where $x_{e,j} > 0$ for all $j=1,\dots,J$ and all $e=1,2$. 
The line integral becomes a summation:
\begin{equation}\label{eq:linCT}
	\int_{\R}\mu_e(\*\nu_i(l)) \, \ud l = \sum_{j=1}^{J} x_{e,j} \int_{\R}b_j(\*\nu_i(l))\, \ud l = \sum_{j=1}^{J} a_{i,j} x_{e,j}   
\end{equation}
where $\*\nu_i(l)  = \*s_i + l  \vec{\*\epsilon}_i \in \R^2$ is a parametrisation of the $i$-th ray emitted from the source $\*s_i$ with direction $\vec{\*\epsilon}_i$, $a_{i,j} \triangleq \int_{\R}b_j(\*\nu_i(l))\,\ud l$ is the contribution of the $j$-th pixel to the $i$-th ray. 
The system matrix $\*A$ is constructed as an under-determined matrix of dimensions $I\times J$ where $I = N_\mathrm{d}\times N_{\theta}$ with $N_\mathrm{d}$ and $N_{\theta}$ being respectively the number of detectors and $N_{\theta}$ and the number of angles (projections), and is defined as $[\*A]_{i,j} = a_{i,j}$, $\forall\; i=1,\dots,I,\;\forall\; j=1,\dots,J$. 
The spectral X-ray mathematical discrete model is based on the Beer's law which provides the X-ray intensity after transmission. The expected number of detected photons $\bar{y}_{i,e}$ is then redefined as a function of the discrete image $\*x_e$ as
\begin{equation}\label{eq:mean_Poisson_DECT}
	\bar{y}_{i,e} (\*x_e) = S_e \+e^{ - [\*A \*x_e]_i } + \eta_{e,i} 
\end{equation}
where $\*x_e = [x_{e,1},\ldots, x_{e,J}]\transp\in\R^J$ is the vector of attenuation coefficients at source energy $e$, $S_e$ is the mean photons flux at the $e$-th energy bin, as we assume a mono-energetic intensity, and $\eta_{e,i}\in\R^+$ is a known additive term representing the expected number of background events (primarily from
scatter). In the case of normal exposure, the number of detected photons follows a Poisson distribution, i.e.,
\begin{equation}\label{eq:poisson_measurment}
	y_{i,e} \sim \+{Poisson}(\bar{y}_{i,e} (\*x_e)) 
\end{equation} 

\noindent and the measurements at each energy bin $e=1,2$ are stored in a vector $\*y_e = [y_{e,1},\dots,y_{e,N_\mathrm{d} \cdot N_{\theta}}]\transp$. 

Although monochromatic X-ray source does not usually hold for scanners in clinical practice, a common effective strategy consists of applying a polychromatic-to-monochromatic source correction pre-processing step \cite{Whiting2006}, and in the rest of the paper we will therefore assume that we have a monoenergetic source or that it has already been appropriately corrected.

\subsection{Low-Dose CT Reconstruction}

In case of low X-ray dose, since the photons counts can be very limited, the Gaussian approximation is no longer applicable as the  logarithm of the data cannot be computed. We therefore chose to perform sparse view \ac{CT} reconstruction from the raw measurements $(\*y_1,\*y_2)$ by solving the minimization problem~\eref{eq:MPML}, with positivity constraints on $(\*x_1,\*x_2)$,  using the Poisson \ac{NLL} functions $L_1$ and $L_2$  defined as
\begin{equation}\label{eq:NLL}
	-L_e(\*x_e,\*y_e) = \sum_{i=1}^{I} y_{e,i}\log \bar{y}_{i,e} (\*x_e) - \bar{y}_{i,e} (\*x_e) , \quad e=1,2
\end{equation}
and the trained regularizer $R_\+{mc}$ derived from the learned filters $(\*D_1^\star,\*D_2^\star)$ as in \eref{eq:pen_MCAOL}. 
Therefore, substituting \eref{eq:NLL} and \eref{eq:pen_MCAOL} into the minimization \eref{eq:MPML}, 
we obtain the following explicit expression for the \ac{MCAOL} \ac{DECT} reconstruction problem:
\begin{eqnarray}\label{eq:NLL_MCAOLPoisson}
	(\*x^{\star}_1,\*x^{\star}_2) = \argmin_{\*x_1,\*x_2 \geq 0} \, \sum_{e=1}^2\rho_e\underbrace{ \sum_{i=1}^{I} y_{e,i}\log \bar{y}_{i,e} (\*x_e) - \bar{y}_{i,e} (\*x_e)}_{ L_e(\*x_e,\*y_e)}  \nonumber \\
	+ \underbrace{\min_{\{\*z_{1,k}\} \atop \{\*z_{2,k}\}} \sum_{k=1}^{K}  \Bigg\{\sum_{e=1}^2 \frac{\gamma_e}{2} \left\| \*d^{\star}_{e,k} \circledast \*x_{e}  - \*z_{e,k}\right\|_2^2 \Bigg\} + \left\| (\*z_{1,k}, \*z_{2,k}) \right\|_{1,0}}_{R_\mathrm{mc}(\*x_1,\*x_2)}
\end{eqnarray} 

We solve the minimization problem (\ref{eq:NLL_MCAOLPoisson}) by the alternating estimation of the sparse feature images and the linear attenuation images $\{\*x_e: e = 1,2\}$. Given the current estimates of the sparse coefficients $\{\*z^t_k: k=1,\ldots, K\}$, the image update $\*x_e^t$ at iteration $t$ is obtained through the following  minimization problem
\begin{eqnarray}
	\*x_e^{t}  =   \argmin_{\*x_e \in (\R^+)^J}  \Phi^t_e(\*x_e) \label{eq:MCAOLoptimization} \\
	\mathrm{with} \;\; \Phi_e(\*x_e) =   \rho_e L_e( \*x_e,\*y_e  )  + \frac{\gamma_e}{2} \sum_{k=1}^{K}  \left\| \*d^{\star}_{e,k} \circledast \*x_e  - \*z^{t}_{e,k}\right\|_2^2   \, .\nonumber
\end{eqnarray}

In this work, we utilized a \ac{LBFGS} algorithm \cite[Chapter 7]{NumOpti2006} to solve \eref{eq:MCAOLoptimization}. The \ac{LBFGS} iterative solver estimates $\*x_e^{t}$ starting the previous iterate $\*x_e^{t-1}$. We define the first estimate as $\*x_e^{t,(0)} = \*x_e^{t}$. At inner iteration $q$, given a current estimate $\*x_e^{t,(q)}$, the new estimate $\*x_e^{t,(q+1)}$ is obtained as
\begin{eqnarray}\label{eq:LBFGS}
	\*x_e^{t,(q+1)} & = & \*x_e^{t,(q)} - s^{\star}\*B^{t,(q)}\nabla\Phi_e^t(\*x_e^{t,(q)}) \nonumber \\
	\mathrm{with}\quad s^{\star} & = & \argmax_{s\in[0,1]}\chi(s) \label{eq:line_search} \\
	\mathrm{and}\quad \chi(s) & = & \Phi_e^t\left(\*x_e^{t,(q)} - s\*B^{t,(q)}\nabla\Phi_e^t(\*x_e^{t,(q)})\right) \nonumber
\end{eqnarray}
where $\*B^{t,(q)}$ is an approximate inverse Hessian of $\Phi_e^t$ evaluated at $\*x_e^{t,(q)}$. The matrix/vector product $\*B^{t,(q)}\nabla\Phi_e^t(\*x_e^{t,(q)})$ in (\ref{eq:LBFGS}) is directly computed (without storing $\*B^{t,(q)}$) from the $m$ previous iterates $\*x_e^{t,(q-p)}$, $p = 0, \ldots, m-1$. An approximate solution of the line-search sub-problem is obtained by backtracking to match the \emph{Wolfe Conditions}. The iterative scheme (i.e., w.r.t. $q$) is repeated until either a convergence criterion is met. A more detailed explanation can be found in \citeasnoun{bousse2020hypo}. We utilized the implementation proposed in \citeasnoun{zhu1997algorithm}. We also used the \ac{LBFGS} algorithm to minimize $L_1(\cdot,\*y_1)$ and $L_2(\cdot,\*y_2)$ (without penalty) in order to obtain initial images $\*x_1^0$ and $\*x_2^0$.

The other part of the alternating scheme is to update the sparse features update $\*z^t_{e,k}$ given the current estimate of $\*x^t_e$ the sparse features update $\*z^t_{e,k}$. This step is achieved using the multi-channel thresholding operator defined in \eref{eq:mthres}. 

\noindent The pseudo-code for \ac{MCAOL} reconstruction algorithm is detailed in Algorithm~\ref{algo:MCAOL Recon}.

\begin{algorithm}[!t]
	\SetAlgoLined
	\KwIn{Initial images $(\*x_1^0, \*x_2^0)$, DECT learned filters $\*D^{\star} = ( \*D_1^{\star}, \*D_2^{\star} )$, joint sparsity weight $\bm\gamma = (\gamma_1,\gamma_2)$, penalty weights $\bm\rho = (\rho_1,\rho_2)$, DE sinogram $\*y = (\*y_1,\*y_2)$, system matrix $\*A$, intensities $(S_1,S_2)$ , \#outer iterations $N_\+{outer}$. }
	\KwOut{Reconstructed images $( \*x^{\star}_1, \*x^{\star}_2 )$}
	\For{$t = 0, \ldots, N_\+{outer}-1$}{
		\textit{Update sparse codes (in parallel)} \;
		\For{k,\dots,K}{
			$(\*z^{t+1}_{1,k}, \*z^{t+1}_{2,k}) \leftarrow$ $\mathcal{T}_{\+{mc}}(\*d^\star_{1,k} \circledast \*x_1^t, \*d^\star_{2,k} \circledast \*x_2^t, \bm\gamma)$ \;
		}
		\textit{Update linear attenuation images} \;
		$\*x^{t+1}_{1} \leftarrow \mathrm{L}\mbox{-}\mathrm{BFGS} (\Phi_1^t , \mathrm{init}= \*x_1^{t} \mid \*y_1, \*z^{t+1}_{1}, \*D_1^{\star}, \*A,S_1,\rho_1,\gamma_1 )  $  \;
		$\*x^{t+1}_{2} \leftarrow \mathrm{L}\mbox{-}\mathrm{BFGS} (\Phi_2^t , \mathrm{init}= \*x_2^{t} \mid \*y_2, \*z^{t+1}_{2}, \*D_1^{\star}, \*A,S_2,\rho_2, ,\gamma_2)  $  \;
	}
	$\*x^\star_{1} \leftarrow \*x^{N_\+{outer}}_{1} $ \; 
	$\*x^\star_{2} \leftarrow \*x^{N_\+{outer}}_{2} $ \; 
	\caption{\ac{MCAOL} Reconstruction Algorithm}\label{algo:MCAOL Recon}
\end{algorithm}

%% file: results.tex
\section{Validation}\label{sec:results}

We validated the proposed methods on two different \ac{DECT} low-dose acquisition setup. In particular, we analyzed  the case of sparse-view \ac{DECT} reconstruction with normal photon dose and the case of extreme low-photon counts with increased number of views. By approximating the dose as the product of the number of views and photon counts, the latter case represents a more challenging scenario since the overall dose considered is lower than the sparse-view case. Our implementation was based on \ac{CONVOLT} \cite{convolt}.

\subsection{Methods Used for Comparison}

The objective of the simulations with sparse views and normal X-ray source intensity is to demonstrate that \ac{MCAOL} achieves improved accuracy compared reconstructing each energy separately by solving \eref{eq:PML} with the \ac{CAOL} regularizer defined in \eref{eq:pen_CAOL}  and with the edge-preserving \ac{TV} regularizers, as well as simultaneously by solving \eref{eq:MPML} with the \ac{JTV} regularizer, respectfully defined as 
\begin{equation}\label{eq:tv} 
	R_\+{tv}  (\*x)  = \sum_{j=1}^J \sum_{k\in\-N_j}  \omega_{j,k} \sqrt{ (x_j - x_k)^2  +  \varepsilon } 
\end{equation}
and 
\begin{equation}	\label{eq:jtv}
	R_\+{jtv}(\*x_1,\*x_2)   = 
	\sum_{j=1}^J \sum_{k\in\-N_j} \omega_{j,k}  \sqrt{(x_{1,j} - x_{1,k})^2  +  (x_{2,j} - x_{2,k})^2  + \varepsilon } 
\end{equation}
where $\-N_j$ denotes the 8 nearest neighboring pixels of pixel $j$ and $\omega_{j,k}$ are weights ($\omega_{j,k}=1$ for axial neighbors and $\omega_{j,k}=1/\sqrt{2}$ for diagonal neighbors), and $\varepsilon>0$ is a small real value to ensure differentiability. For each method, we used the \ac{LBFGS} solver to estimate $\*x_1$ and $\*x_2$. 

The experiment with extreme low-counts aims at demonstrating that considering a weighted least-squares approximation of the log-likelihood function no longer guarantees effective reconstruction results, instead the exact Poisson statistics should be accounted. This results in a degradation of the performance of \ac{CAOL} when optimized through the \ac{PWLS} solver while using the quasi-Newton solver \ac{LBFGS} leads to improved qualitative and quantitative results.

\subsection{Methodology}

All experiments were validated by generating the \ac{DECT} measurements as in Eq.~\eref{eq:mean_Poisson_DECT} and then running $n=20$ Poisson noise instances as in Eq.~\eref{eq:poisson_measurment} from a \ac{GT} image $\*x_e^\star = [x_{e,1}^*,\dots,x_{e,J}^*]\transp\in\R^J$, $e=1,2$. As performance metrics, we considered the mean absolute bias ($\+{AbsBias}$) error function defined as follows
\begin{equation}
	\+{AbsBias} = \frac{1}{N_{\-R}}\frac{1}{N_{\+{noise}}} \sum_{j\in\-R} \sum_{n=1}^{N_{\mathrm{noise}}}\left| x_{e,j}^{[n]} - x_{e,j}^*  \right|
\end{equation}
where $x_{e,j}^{[n]}$ indicates the reconstructed linear attenuation coefficient at image pixel $j$ from the $n$-th Poisson noise replicate, $\-R$ is the spatial region of interest and $N_{\-R}$ is the number of pixels in the region $\-R$. Furthermore, we compute the \ac{STD} defined as
\begin{equation}
	\+{STD} = \frac{1}{N_{\-R}} \sum_{j\in\-R} \sqrt{\frac{1}{N_{\+{noise}}} \sum_{n=1}^{N_{\mathrm{noise}}}\left( x_{e,j}^{[n]} - \bar{x}_{e,j}  \right)^2}
\end{equation}
where 
$\bar{x}_{e,j} = \frac{1}{N_{\+{noise}}}\sum_{n=1}^{N_{\mathrm{noise}}} x_{e,j}^{[n]}$. In this work $\-R$ corresponds to the nonnegative pixels region of $\*x_e^\star$ and is the same for both energy levels. 

The simulations were repeated for all the methods, for different values of the regularization parameters in the objective functions \eref{eq:PML} and \eref{eq:MPML} in order to plot $\+{AbsBias}$ / $\+{STD}$ curves (Fig.~\ref{fig:Bias_vs_std_xcat}, \ref{fig:Bias_vs_std_clin} and \ref{fig:Bias_vs_std_clin_ld}). The quality of the reconstruction is assessed by the proximity of the curve to the origin. Training and reconstruction were performed according to the below-described settings.

\paragraph{Training}

The optimization problem (\ref{eq:dual_d_opt}) is minimized using the \ac{BPEGM} algorithm \cite{Chun2019} with normalized input dataset.  To investigate the trade-off between accuracy and features sparsity, we tested (\ref{eq:dual_d_opt}) with different values of $\gamma_1 = \gamma_2$ with filter $(\*d_{1,k}, \*d_{2,k})$ of dimension $P = 49$ and number of filters $K=49$. For each simulation, we tuned $\gamma_1 = \gamma_2$ by testing different values to investigate the effect. For all the datasets, we have used a training of $L=25$ images for each energy $e$. Regarding the \ac{BPEGM} algorithm, we set the tolerance value equal to $10^{-4}$ and the maximum number of iterations to $3\times 10^3$. For the \ac{CAOL} training algorithm, we have used the same settings as detailed for \ac{MCAOL} except that we tuned a single regularization weight $\alpha$ in the optimization problem \eref{eq:cost_tr_CAOL} for each separate energy channel. 

\paragraph{Reconstruction} \Ac{MCAOL} and \ac{JTV} reconstructions were achieved by solving \eref{eq:MPML} with $R_{\+{mc}}$ defined as \eref{eq:pen_MCAOL} and \eref{eq:jtv} respectively, while \ac{CAOL} and \ac{TV} reconstructions by solving \eref{eq:PML} for each energy bin $e=1,2$ separately with $R$ defined as \eref{eq:pen_CAOL} and \eref{eq:tv} respectively. \Ac{MCAOL} and \ac{CAOL} were achieved using $N_{\+{outer}}=300$ outer iterations while the inner image update is obtained using the \ac{LBFGS} algorithm with $300$ iterations. The $(\gamma_1,\gamma_2)$-values and $\beta$-values were the same as for training. \Ac{TV} and \ac{JTV} reconstructions were achieved with the \ac{LBFGS} algorithm with $300$ iterations. The measurements were obtained from the \ac{GT} images $\*x_e^\star$ outside of the training set and the reconstructions were repeated for each noise instance $n$, for a range of  $(\rho_1,\rho_2)$-values with $\rho_1 = \rho_2$ and for a range of $\beta$-values, in order to obtain $\+{AbsBias}$-versus-$\+{STD}$ curves.

We performed sparse-views and low-dose experiments on a simulated \ac{XCAT} phantom and clinical data to assess the potential of the method for medical practice as detailed below. 
The experiments were conducted with fixed X-ray dose amount, i.e., by selecting the number of angles and the X-ray source intensity, and we evaluated the quality of the linear attenuation  images reconstructed with different methods, both qualitatively and quantitatively.

\subsection{Results on XCAT Phantom}

For the unsupervised \ac{MCAOL} and \ac{CAOL} training, the numerical data consists of 1-mm pixel-width $512\times 512$ torso axial slice images generated from the \ac{XCAT} phantom for $60$ keV and $120$ keV energies. 

We utilized 20 slice pairs from the \ac{XCAT} phantom, each pair consisting of a slice at $E_1=60$~KeV and a slice at $E_2=120$~KeV, to train the filters. An additional slice pair---not part of the training dataset---was used to generate the projection data  (i.e, following \eref{eq:mean_Poisson_DECT} and \eref{eq:poisson_measurment}) for reconstruction, as detailed below. We used the \ac{MCAOL} weights parameters $\gamma_1 = \gamma_2 = 800$ and the \ac{CAOL} parameter $\alpha = 0.01$. 

\begin{figure}[!h]
	\centering
	\subcaptionbox{Learned $l_{1,0}$ filters via \ac{MCAOL} \label{subfig:MCAOL:filters:XCAT_l0_l10}}
	{
		\begin{tabular}{cc}
			\includegraphics[scale=0.2, clip]{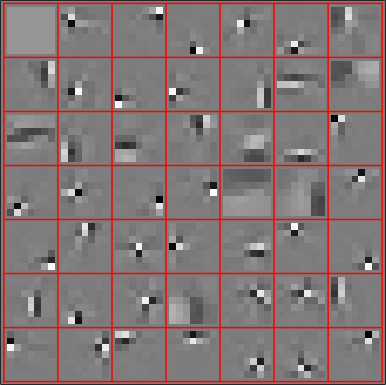} &
			\includegraphics[scale=0.2, clip]{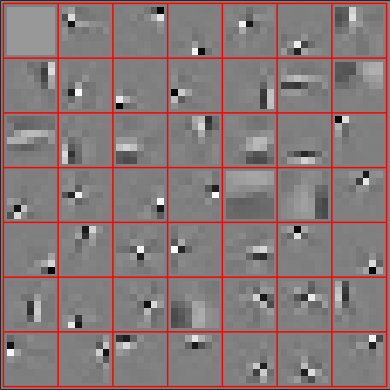} \\
			{\small ($E_1$), $\gamma_1 = 800$} & {\small ($E_2$), $\gamma_2 = 800$} 
		\end{tabular}
	}
	\hspace{0.1\linewidth}
	\subcaptionbox{Learned $l_0$ filters via \ac{CAOL} \label{subfig:MCAOL:filters:XCAT_l0_l0}}
	{
		\begin{tabular}{cc}
			\includegraphics[scale=0.2, clip]{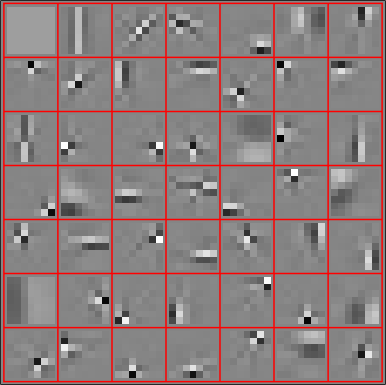} &
			\includegraphics[scale=0.2, clip]{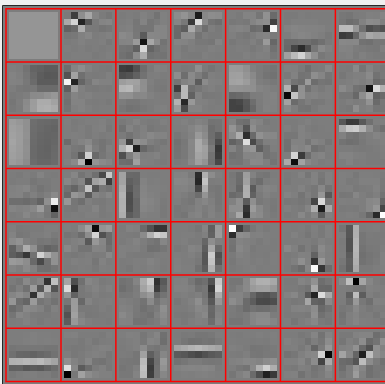} \\
			{\small ($E_1$), $\alpha = 0.01$} & {\small ($E_2$), $\alpha = 0.01$} 
		\end{tabular}
	}
	\caption{
		Learned filters $\{(\*d_{1,k}, \*d_{2,k})\}$ with $P=K=49$ using the \ac{XCAT} training dataset, for \subref{subfig:MCAOL:filters:XCAT_l0_l10} \ac{MCAOL} and \subref{subfig:MCAOL:filters:XCAT_l0_l0} \ac{CAOL}. 
	} \label{fig:MCAOL:filters:XCAT_l0}
\end{figure}

\begin{figure}[!h]
	\centering
	\subcaptionbox{ $\bm{z}_{1,k}$, CAOL \label{subfig:z1_sep_xcat}}
	{ 
		\includegraphics[width=0.235\linewidth]{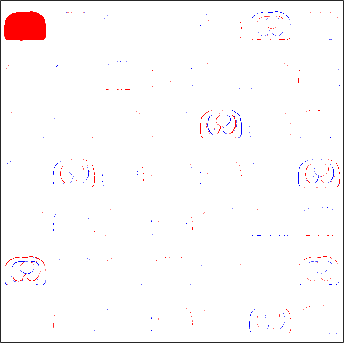}
	}
	\hspace{-0.025\linewidth}
	\subcaptionbox{ $\bm{z}_{2,k}$, CAOL \label{subfig:z2_sep_xcat}}
	{ 
		\includegraphics[width=0.235\linewidth]{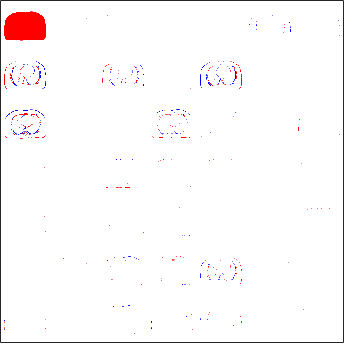}
	}
	\hspace{-0.01\linewidth}
	\subcaptionbox{ $\bm{z}_{1,k}$, MCAOL  \label{subfig:z1_joint_xcat}}
	{ 
		\includegraphics[width=0.235\linewidth]{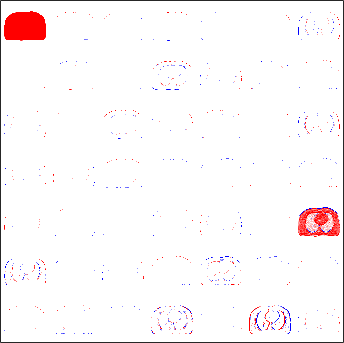}
	}
	\hspace{-0.025\linewidth}
	\subcaptionbox{ $\bm{z}_{2,k}$, MCAOL \label{subfig:z2_joint_xcat}}
	{ 
		\includegraphics[width=0.235\linewidth]{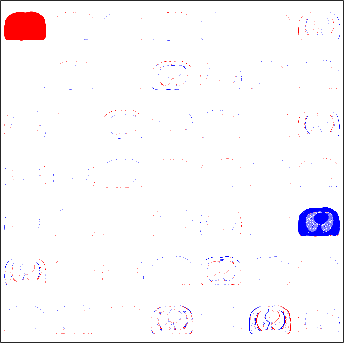}
	}
	
	\caption{\ac{XCAT} Phantom: estimated sparse feature maps $\bm{z}_{e,k}$ for $e=1,2$ and $k=1,\dots,49$ using \ac{CAOL} (\subref{subfig:z1_sep_xcat} and \subref{subfig:z2_sep_xcat}) and \ac{MCAOL} (\subref{subfig:z1_joint_xcat} and \subref{subfig:z2_joint_xcat}); color scale: red for positive values, blue for negative values.}\label{fig:z_xcat}

\end{figure}

\begingroup
\tabcolsep = 8.2pt
\renewcommand\arraystretch{0.1} 
\setlength{\tabcolsep}{8pt}
\begin{figure}[!h]
	\centering
	\subcaptionbox{\Ac{XCAT} \label{sugfig:xcat_gt}}
	{
		\begin{tabular}{c}
			\begin{tikzpicture}
				\begin{scope}[spy using outlines={rectangle,yellow,magnification=1.25,size=12mm,connect spies}]
					\node {\includegraphics[viewport=5 5 295 295, clip, width=2.3cm, height=2.3cm]{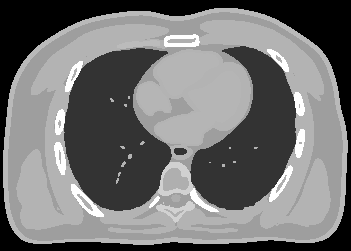}};
					\spy on (0.3,-0.65) in node [left] at (-0.3,-1.2);   
				\end{scope}
			\end{tikzpicture} \\
			\begin{tikzpicture}
				\begin{scope}[spy using outlines={rectangle,yellow,magnification=1.25,size=12mm,connect spies}]
					\node {\includegraphics[viewport=5 5 295 295, clip, width=2.3cm, height=2.3cm]{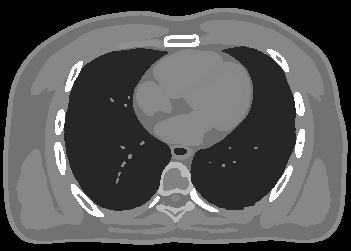}};
					\spy on (0.3,-0.65) in node [left] at (-0.3,-1.2);
				\end{scope}
			\end{tikzpicture}
		\end{tabular}	
	}
	\hspace{-0.095\linewidth}
	\subcaptionbox{No prior \label{sugfig:xcat_lbfgs}}
	{
		\begin{tabular}{c}
			\begin{tikzpicture}
				\begin{scope}[spy using outlines={rectangle,yellow,magnification=1.25,size=12mm,connect spies}]
					\node {\includegraphics[viewport=5 5 295 295, clip, width=2.3cm, height=2.3cm]{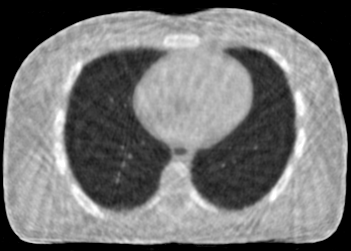}};
					\spy on (0.3,-0.65) in node [left] at (-0.3,-1.2); 
				\end{scope}
			\end{tikzpicture} \\
			\begin{tikzpicture}
				\begin{scope}[spy using outlines={rectangle,yellow,magnification=1.25,size=12mm,connect spies}]
					\node {\includegraphics[viewport=5 5 295 295, clip, width=2.3cm, height=2.3cm]{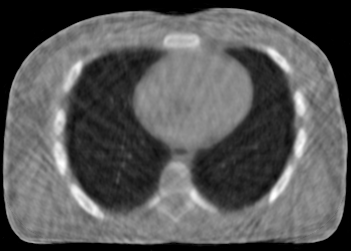}};
					\spy on (0.3,-0.65) in node [left] at (-0.3,-1.2);
				\end{scope}
			\end{tikzpicture}
		\end{tabular}	
	}
	\hspace{-0.095\linewidth}
	\subcaptionbox{\Ac{CAOL} \label{sugfig:xcat_caol}}
	{
		\begin{tabular}{c}
			\begin{tikzpicture}
				\begin{scope}[spy using outlines={rectangle,yellow,magnification=1.25,size=12mm,connect spies}]
					\node {\includegraphics[viewport=5 5 295 295, clip, width=2.3cm, height=2.3cm]{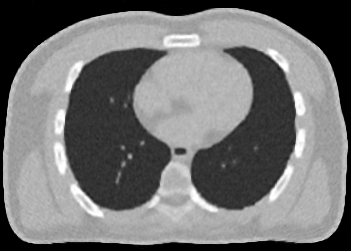}};
					\spy on (0.3,-0.65) in node [left] at (-0.3,-1.2);
				\end{scope}
			\end{tikzpicture} \\
			\begin{tikzpicture}
				\begin{scope}[spy using outlines={rectangle,yellow,magnification=1.25,size=12mm,connect spies}]
					\node {\includegraphics[viewport=5 5 295 295, clip, width=2.3cm, height=2.3cm]{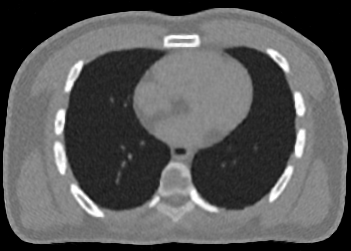}};
					\spy on (0.3,-0.65) in node [left] at (-0.3,-1.2);
				\end{scope}
			\end{tikzpicture}
		\end{tabular}	
	}
	\hspace{-0.095\linewidth}
	\subcaptionbox{\Ac{MCAOL}\label{sugfig:xcat_mcaol} }
	{
		\begin{tabular}{c}
			\begin{tikzpicture}
				\begin{scope}[spy using outlines={rectangle,yellow,magnification=1.25,size=12mm,connect spies}]
					\node {\includegraphics[viewport=5 5 295 295, clip, width=2.3cm, height=2.3cm]{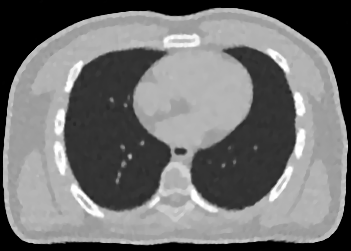}};
					\spy on (0.3,-0.65) in node [left] at (-0.3,-1.2);
				\end{scope}
			\end{tikzpicture} \\
			\begin{tikzpicture}
				\begin{scope}[spy using outlines={rectangle,yellow,magnification=1.25,size=12mm,connect spies}]
					\node {\includegraphics[viewport=5 5 295 295, clip, width=2.3cm, height=2.3cm]{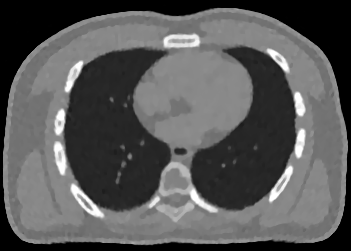}};
					\spy on (0.3,-0.65) in node [left] at (-0.3,-1.2);
				\end{scope}
			\end{tikzpicture}
		\end{tabular}	
	}
	\hspace{-0.095\linewidth}
	\subcaptionbox{\Ac{TV} \label{sugfig:xcat_tv}}
	{
		\begin{tabular}{c}
			\begin{tikzpicture}
				\begin{scope}[spy using outlines={rectangle,yellow,magnification=1.25,size=12mm,connect spies}]
					\node {\includegraphics[viewport=5 5 295 295, clip, width=2.3cm, height=2.3cm]{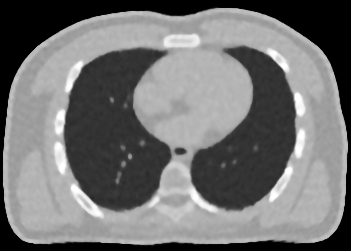}};
					\spy on (0.3,-0.65) in node [left] at (-0.3,-1.2);
				\end{scope}
			\end{tikzpicture} \\
			\begin{tikzpicture}
				\begin{scope}[spy using outlines={rectangle,yellow,magnification=1.25,size=12mm,connect spies}]
					\node {\includegraphics[viewport=5 5 295 295, clip, width=2.3cm, height=2.3cm]{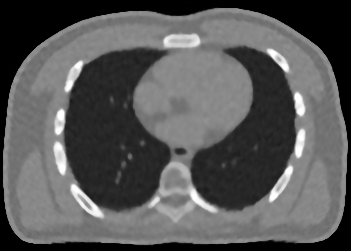}};
					\spy on (0.3,-0.65) in node [left] at (-0.3,-1.2);
				\end{scope}
			\end{tikzpicture}
		\end{tabular}	
	}
	\hspace{-0.095\linewidth}
	\subcaptionbox{\Ac{JTV} \label{sugfig:xcat_jtv}}
	{
		\begin{tabular}{c}
			\begin{tikzpicture}
				\begin{scope}[spy using outlines={rectangle,yellow,magnification=1.25,size=12mm,connect spies}]
					\node {\includegraphics[viewport=5 5 295 295, clip, width=2.3cm, height=2.3cm]{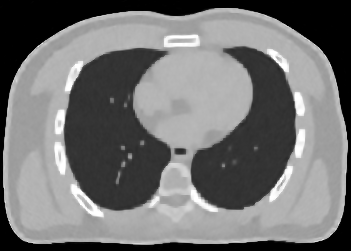}};
					\spy on (0.3,-0.65) in node [left] at (-0.3,-1.2);
				\end{scope}
			\end{tikzpicture} \\
			\begin{tikzpicture}
				\begin{scope}[spy using outlines={rectangle,yellow,magnification=1.25,size=12mm,connect spies}]
					\node {\includegraphics[viewport=5 5 295 295, clip, width=2.3cm, height=2.3cm]{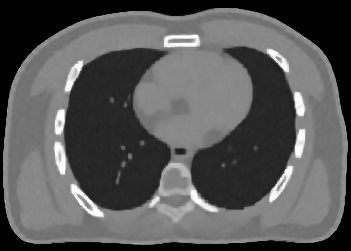}};
					\spy on (0.3,-0.65) in node [left] at (-0.3,-1.2);
				\end{scope}
			\end{tikzpicture}
		\end{tabular}	
	}
	\caption{
		Comparison of reconstructed \ac{XCAT} phantom from different reconstruction methods for sparse-view \ac{CT} with top row corresponding to high energy $E_1 = 120$ keV and bottom row to low energy $E_2 = 60$ keV: \subref{sugfig:xcat_gt} \ac{GT} \ac{XCAT} test image, \subref{sugfig:xcat_lbfgs} minimization of the \ac{NLL} function without prior, \subref{sugfig:xcat_caol} \ac{CAOL} reconstruction, \subref{sugfig:xcat_mcaol} \ac{MCAOL} joint reconstruction, \subref{sugfig:xcat_tv} separate reconstruction using \ac{TV} prior and \subref{sugfig:xcat_jtv} joint reconstruction using \ac{JTV} prior. 
	}
	\label{fig:CTreconXCAT}
	
\end{figure}		
\endgroup

Fig.~\ref{fig:MCAOL:filters:XCAT_l0} shows the pairs $(\*d_{1,k}, \*d_{2,k})$ of learned convolutional filters obtained by \ac{MCAOL} (Fig.~\ref{subfig:MCAOL:filters:XCAT_l0_l10}) and separate learning with \ac{CAOL} (Fig.~\ref{subfig:MCAOL:filters:XCAT_l0_l0}). From a qualitative point of view, we observe that \ac{MCAOL} filter pairs $(\*d_{1}, \*d_{2})$ look similar as the edges are identical in the $2$ energy images,  as opposed to the \ac{CAOL} filters. 

\begin{figure}[!h]
	\centering
	\subcaptionbox{$60$ keV \label{subfig:Bias_vs_std_xcat_e1}}
	{
		\begin{tikzpicture}[scale=.75] 
			\begin{axis}[
				xlabel={$\+{STD}$},
				ylabel={$\+{AbsBias}$},
				grid = major,
				legend columns=2,
				legend cell align=left,
				legend entries={MCAOL, CAOL, TV, JTV},
				legend style={at={(0.7,0.2)},anchor=north}
				]				
				\addplot[color=blue,mark=*] table[x=xMCAOL, y=yMCAOL] {Fig_5a.txt};
				\addplot[color=black,mark=square*] table[x=xCAOL, y=yCAOL] {Fig_5a.txt};
				\addplot[color=red,mark=diamond*] table[x=xTV, y=yTV] {Fig_5a.txt};
				\addplot[color=green,mark=pentagon*] table[x=xJTV, y=yJTV] {Fig_5a.txt};				
			\end{axis}
		\end{tikzpicture}
	}
	\subcaptionbox{$120$ keV \label{subfig:Bias_vs_std_xcat_e2}}
	{
		\begin{tikzpicture}[scale=.75] 
			\begin{axis}[
				xlabel={$\+{STD}$},
				ylabel={$\+{AbsBias}$},
				grid = major,
				legend columns=2,
				legend cell align=left,
				legend entries={MCAOL, CAOL, TV, JTV},
				legend style={at={(0.7,0.19)},anchor=north}
				]
				\addplot[color=blue,mark=*] table[x=xMCAOL, y=yMCAOL] {Fig_5b.txt};
				\addplot[color=black,mark=square*] table[x=xCAOL, y=yCAOL] {Fig_5b.txt};
				\addplot[color=red,mark=diamond*] table[x=xTV, y=yTV] {Fig_5b.txt};
				\addplot[color=green,mark=pentagon*] table[x=xJTV, y=yJTV] {Fig_5b.txt};
			\end{axis}
		\end{tikzpicture}
	}
	\caption{Plot of the mean absolute bias ($\+{AbsBias}$) versus the standard deviation ($\+{STD}$) for the \ac{XCAT} phantom at \subref{subfig:Bias_vs_std_xcat_e1} low X-ray source energy ($60$~keV) and \subref{subfig:Bias_vs_std_xcat_e2} high X-ray source energy ($120$ keV).} \label{fig:Bias_vs_std_xcat}
	
\end{figure}
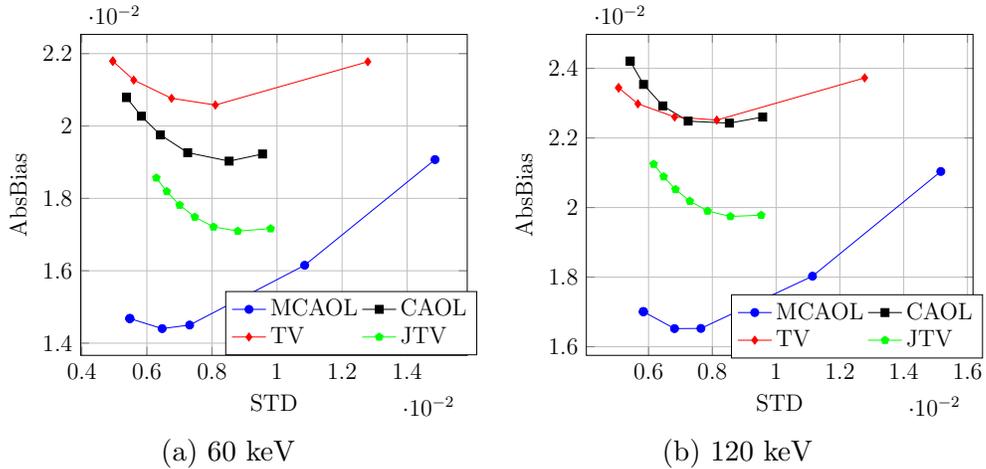

In order to generate the  sparse-view \ac{DECT} projection measurements \eref{eq:mean_Poisson_DECT}, we modeled the projector $\*A$ with a 2-mm \ac{FWHM} resolution parallel beam system and we used 1-mm pixel-width $406\times 406$ \ac{GT} torso axial-slice images with attenuation coefficients $\*x^\star_1, \*x^\star_2$ at energies $120$ keV (high) and $60$ keV (low) which differs from the training examples. The simulation consisted on generating sparse-view sinograms with $406$ detector pixels and $60$ regularly spaced projection angles, where $360^{\circ}$ is the full view rotation. A monochromatic source with $\bar{S}_e = 10^5$ incident photons and $100$ background events was used to generate each sinogram.  

To support the statement that joint sparsity of allows both images to inform each other, which makes the estimation of $\*z_1$ and $\*z_2$ more robust, we show the estimated feature maps in Fig.~\ref{fig:z_xcat} obtained using the \ac{XCAT} data. By comparing the estimated sparse feature maps $\bm{z}_{e,k}$ for $e=1,2$ and $k=1,\dots,49$ for separate reconstructions (Fig.~\ref{subfig:z1_sep_xcat} and Fig.~\ref{subfig:z2_sep_xcat}) and joint reconstruction (Fig.~\ref{subfig:z1_joint_xcat} and Fig.~\ref{subfig:z2_joint_xcat}), there are no similarities between the feature maps obtained from separate reconstructions while the feature maps obtained from joint reconstruction have similar structures.

Fig.~\ref{fig:CTreconXCAT} shows the \ac{XCAT} \ac{GT} and the reconstruction images for both 60 keV and 120 keV energies obtained by \ac{MCAOL} and the other algorithms used for comparison. The images are obtained using the parameters which corresponds to the minimum $\+{AbsBias}$ shown in Figs.~\ref{subfig:Bias_vs_std_xcat_e1} and \ref{subfig:Bias_vs_std_xcat_e2}. It is worth noting that \ac{MCAOL} (Fig.~\ref{sugfig:xcat_mcaol}) manages to substantially reduce the noise as compared with \ac{CAOL} (Fig.~\ref{sugfig:xcat_caol}). 

Fig.~\ref{subfig:Bias_vs_std_xcat_e1} and Fig.~\ref{subfig:Bias_vs_std_xcat_e2} show the $\+{AbsBias}$ against the $\+{STD}$ results respectively for low and high X-ray source energy. Among the methods used for comparison, \ac{TV} promotes sparsity of the gradient, while \ac{JTV} promotes joint sparsity of the $2$ gradients and therefore are particularly well-suited for \ac{XCAT}. Despite this observation, it is possible to show that the minimum $\+{AbsBias}$ obtained by \ac{MCAOL} outperforms all other algorithms, or in other words by fixing the \ac{STD}, the $\+{AbsBias}$ achieved by \ac{MCAOL} is always lower while it is possible to claim that by fixing the $\+{AbsBias}$, the $\+{STD}$ of \ac{MCAOL} is reduced.

\subsection{Results on Clinical Data}\label{subsec:res_clin}

We utilized images reconstructed from data acquired on Philips IQon Spectral CT and reconstructed with a \ac{MBIR} technique \cite{IQon}. 
All patients provided signed permission for the use of their clinical data for scientific purposes and anonymous publication of data. 

The experiment was conducted in a similar fashion as for the \ac{XCAT} simulation. We selected 22 slice pairs from a full body patient scan with $0.902$-mm pixel-width and  $512 \times 512$ image size for the training dataset corresponding to thorax. The energies used in this study are $70$ keV and $140$ keV. An additional slice pair was used to generate the projection data for reconstruction, as detailed below.

\begin{figure}[h!]
	\centering
	\subcaptionbox{Learned $l_{1,0}$ filters via \ac{MCAOL} \label{subfig:MCAOL:filters:XCLIN_l0_l10}}
	{
		\begin{tabular}{cc}
			\includegraphics[scale=0.2, clip]{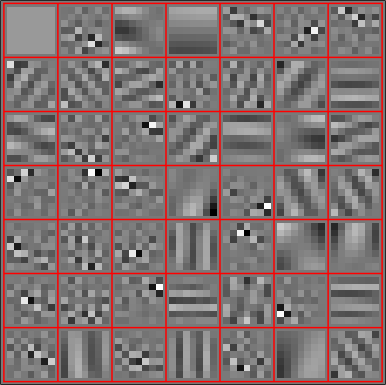} &
			\includegraphics[scale=0.2, clip]{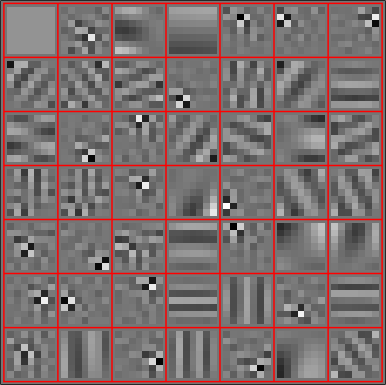} \\
			{\small ($E_1$), $\gamma_1 = 10^4$} & {\small ($E_2$), $\gamma_2 = 10^4$} 
		\end{tabular}
	}
	\hspace{0.1\linewidth}
	\subcaptionbox{Learned $l_0$ filters via \ac{CAOL}  \label{subfig:MCAOL:filters:XCLIN_l0_l0}}
	{
		\begin{tabular}{cc}
			\includegraphics[scale=0.2, clip]{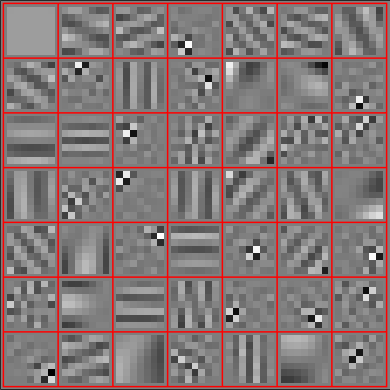} &
			\includegraphics[scale=0.2, clip]{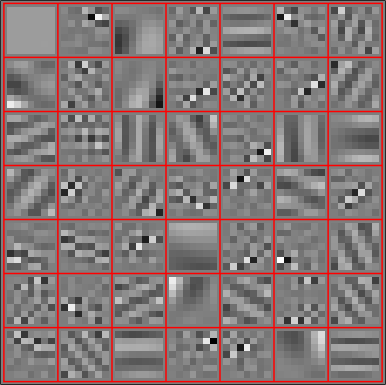} \\
			{\small ($E_1$), $\alpha = 10^{-4}$} & {\small ($E_2$), $\alpha = 10^{-4}$} 
		\end{tabular}
	}
	\caption{
		Learned filters $\{(\*d_{1,k}, \*d_{2,k})\}$ with $P=K=49$ using the clinical training dataset, for \subref{subfig:MCAOL:filters:XCLIN_l0_l10} \ac{MCAOL} and \subref{subfig:MCAOL:filters:XCLIN_l0_l0} \ac{CAOL}.
	} \label{fig:MCAOL:filters:XCLIN_l0}
\end{figure}

The pair of trained filters $(\*d_{1,k}, \*d_{2,k})$ obtained by both \ac{MCAOL} and  \ac{CAOL} unsupervised learning is shown in Fig.~\ref{fig:MCAOL:filters:XCLIN_l0}; we used the parameters $\gamma_1 = \gamma_2 = 10^4$ and the \ac{CAOL} parameter $\alpha = 10^{-4}$. 

To generate the  sparse-view \ac{DECT} measurements \eref{eq:mean_Poisson_DECT}, we used the same geometrical and noise settings as for the XCAT simulation except that we used $451$ detector pixels and $451\times 451$ \ac{GT} thorax images with attenuation coefficients $\*x^\star_1, \*x^\star_2$ at energies $140$ keV (high) and $70$ keV (low) which differs from the training examples. 

\begin{figure}[!h]
	\centering
	\subcaptionbox{ $\bm{z}_{1,k}$, CAOL \label{subfig:z1_sep_clinical}}
	{ 
		\includegraphics[width=0.235\linewidth]{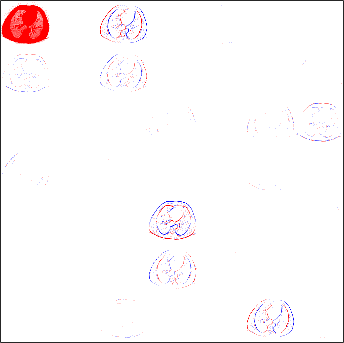}
	}
	\hspace{-0.025\linewidth}
	\subcaptionbox{ $\bm{z}_{2,k}$, CAOL \label{subfig:z2_sep_clinical}}
	{ 
		\includegraphics[width=0.235\linewidth]{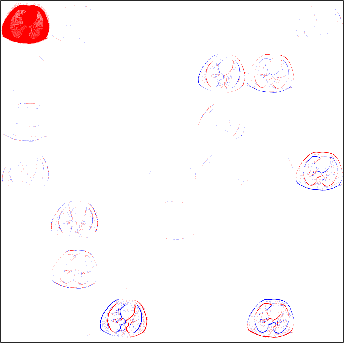}
	}
	\hspace{-0.01\linewidth}
	\subcaptionbox{ $\bm{z}_{1,k}$, MCAOL  \label{subfig:z1_joint_clinical}}
	{ 
		\includegraphics[width=0.235\linewidth]{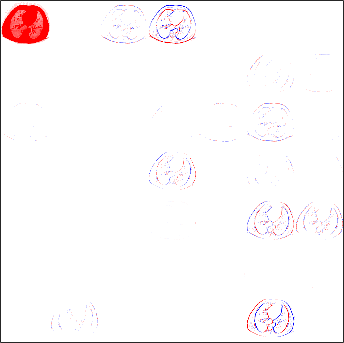}
	}
	\hspace{-0.025\linewidth}
	\subcaptionbox{ $\bm{z}_{2,k}$, MCAOL \label{subfig:z2_joint_clinical}}
	{ 
		\includegraphics[width=0.235\linewidth]{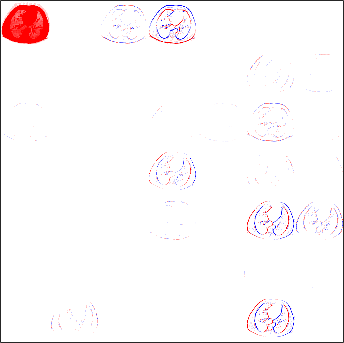}
	}
	
	\caption{Clinical data: estimated sparse feature maps $\bm{z}_{e,k}$ for $e=1,2$ and $k=1,\dots,49$ using \ac{CAOL} (\subref{subfig:z1_sep_clinical} and \subref{subfig:z2_sep_clinical}) and \ac{MCAOL} (\subref{subfig:z1_joint_clinical} and \subref{subfig:z2_joint_clinical}); color scale: red for positive values, blue for negative values.}\label{fig:z_clinical}
	
\end{figure}

In Fig.~\ref{fig:z_clinical} we show the estimated feature maps obtained using the clinical data. As already noted previously with the \ac{XCAT} data, by comparing the estimated sparse feature maps $\bm{z}_{e,k}$ for $e=1,2$ and $k=1,\dots,49$ for separate reconstructions (Fig.~\ref{subfig:z1_sep_clinical} and Fig.~\ref{subfig:z2_sep_clinical}) and joint reconstruction (Fig.~\ref{subfig:z1_joint_clinical} and Fig.~\ref{subfig:z2_joint_clinical}), there are no similarities between the feature maps obtained from separate reconstructions while the feature maps obtained from joint reconstruction have similar structures.

\begingroup
\tabcolsep = 8pt
\def\arraystretch{0} 	

\begin{figure}[!ht]
	\centering
	\subcaptionbox{\centering \Ac{GT} \label{sugfig:clin_gt}}
	{
		\begin{tabular}{c}
			\begin{tikzpicture}
				\begin{scope}[spy using outlines={rectangle,yellow,magnification=1.25,size=10mm,connect spies}]
					\node {\includegraphics[viewport=15 50 295 330, clip, width=2.3cm, height=2.3cm]{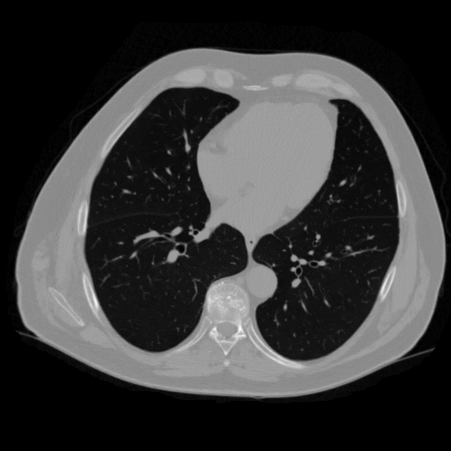}};
					\spy on (0.6,-0.60) in node [left] at (-0.3,-1);   
				\end{scope}
			\end{tikzpicture} \\
			\begin{tikzpicture}
				\begin{scope}[spy using outlines={rectangle,yellow,magnification=1.25,size=10mm,connect spies}]
					\node {\includegraphics[viewport=15 50 295 330, clip, width=2.3cm, height=2.3cm]{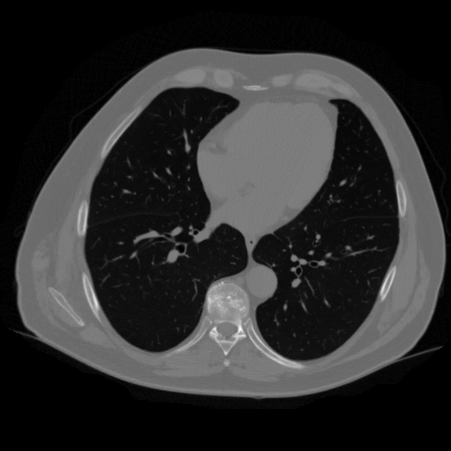}};
					\spy on (0.6,-0.60) in node [left] at (-0.3,-1);  
				\end{scope}
			\end{tikzpicture}
		\end{tabular}	
	}
	\hspace{-0.08\linewidth}
	\subcaptionbox{No prior \label{sugfig:clin_lbfgs}}
	{
		\begin{tabular}{c}
			\begin{tikzpicture}
				\begin{scope}[spy using outlines={rectangle,yellow,magnification=1.25,size=10mm,connect spies}]
					\node {\includegraphics[viewport=15 50 295 330, clip, width=2.3cm, height=2.3cm]{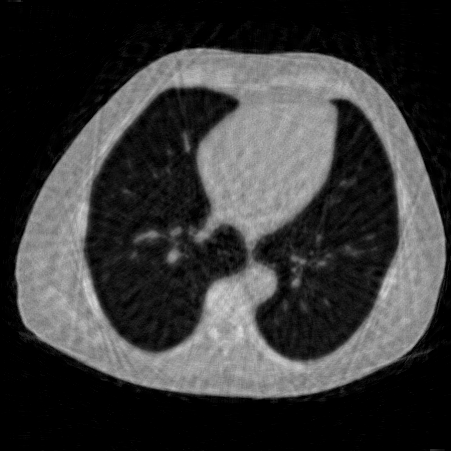}};
					\spy on (0.6,-0.60) in node [left] at (-0.3,-1);  
				\end{scope}
			\end{tikzpicture} \\
			\begin{tikzpicture}
				\begin{scope}[spy using outlines={rectangle,yellow,magnification=1.25,size=10mm,connect spies}]
					\node {\includegraphics[viewport=15 50 295 330, clip, width=2.3cm, height=2.3cm]{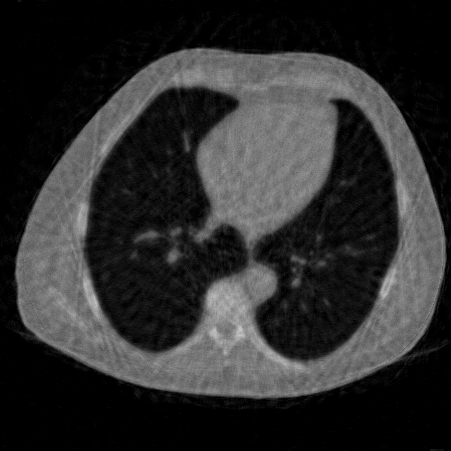}};
					\spy on (0.6,-0.60) in node [left] at (-0.3,-1);   
				\end{scope}
			\end{tikzpicture}
		\end{tabular}	
	}
	\hspace{-0.08\linewidth}
	\subcaptionbox{\Ac{CAOL} \label{sugfig:clin_caol}}
	{
		\begin{tabular}{c}
			\begin{tikzpicture}
				\begin{scope}[spy using outlines={rectangle,yellow,magnification=1.25,size=10mm,connect spies}]
					\node {\includegraphics[viewport=15 50 295 330, clip, width=2.3cm, height=2.3cm]{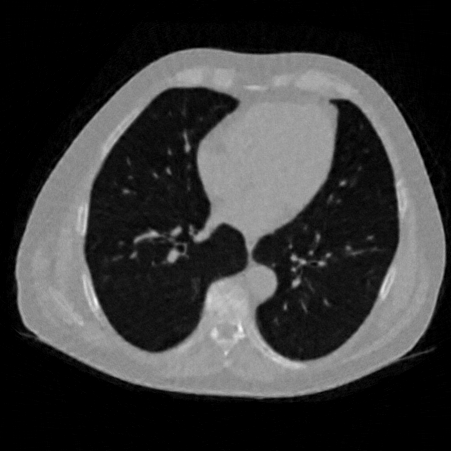}};
					\spy on (0.6,-0.60) in node [left] at (-0.3,-1);   
				\end{scope}
			\end{tikzpicture} \\
			\begin{tikzpicture}
				\begin{scope}[spy using outlines={rectangle,yellow,magnification=1.25,size=10mm,connect spies}]
					\node {\includegraphics[viewport=15 50 295 330, clip, width=2.3cm, height=2.3cm]{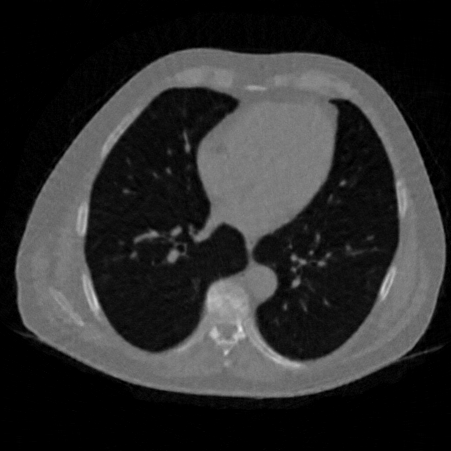}};
					\spy on (0.6,-0.60) in node [left] at (-0.3,-1);  
				\end{scope}
			\end{tikzpicture}
		\end{tabular}	
	}
	\hspace{-0.08\linewidth}
	\subcaptionbox{\Ac{MCAOL} \label{sugfig:clin_mcaol}}
	{
		\begin{tabular}{c}
			\begin{tikzpicture}
				\begin{scope}[spy using outlines={rectangle,yellow,magnification=1.25,size=10mm,connect spies}]
					\node {\includegraphics[viewport=15 50 295 330, clip, width=2.3cm, height=2.3cm]{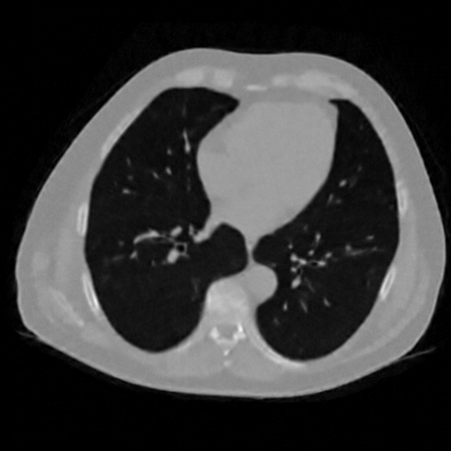}};
					\spy on (0.6,-0.60) in node [left] at (-0.3,-1);  
				\end{scope}
			\end{tikzpicture} \\
			\begin{tikzpicture}
				\begin{scope}[spy using outlines={rectangle,yellow,magnification=1.25,size=10mm,connect spies}]
					\node {\includegraphics[viewport=15 50 295 330, clip, width=2.3cm, height=2.3cm]{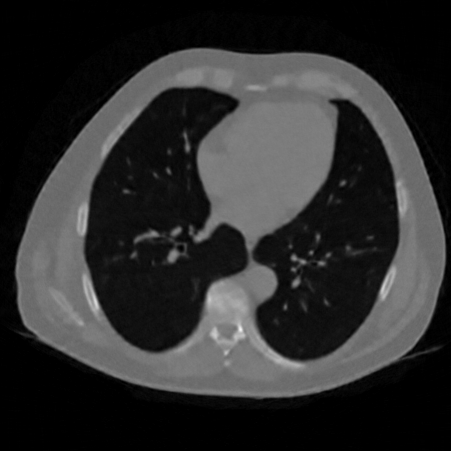}};
					\spy on (0.6,-0.60) in node [left] at (-0.3,-1);   
				\end{scope}
			\end{tikzpicture}
		\end{tabular}	
	}
	\hspace{-0.08\linewidth}
	\subcaptionbox{\Ac{TV} \label{sugfig:clin_tv}}
	{
		\begin{tabular}{c}
			\begin{tikzpicture}
				\begin{scope}[spy using outlines={rectangle,yellow,magnification=1.25,size=10mm,connect spies}]
					\node {\includegraphics[viewport=15 50 295 330, clip, width=2.3cm, height=2.3cm]{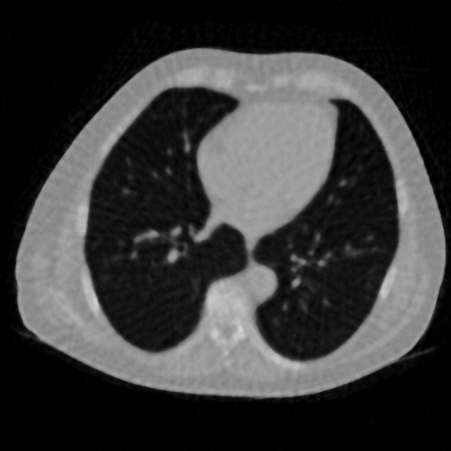}};
					\spy on (0.6,-0.60) in node [left] at (-0.3,-1);   
				\end{scope}
			\end{tikzpicture} \\
			\begin{tikzpicture}
				\begin{scope}[spy using outlines={rectangle,yellow,magnification=1.25,size=10mm,connect spies}]
					\node {\includegraphics[viewport=15 50 295 330, clip, width=2.3cm, height=2.3cm]{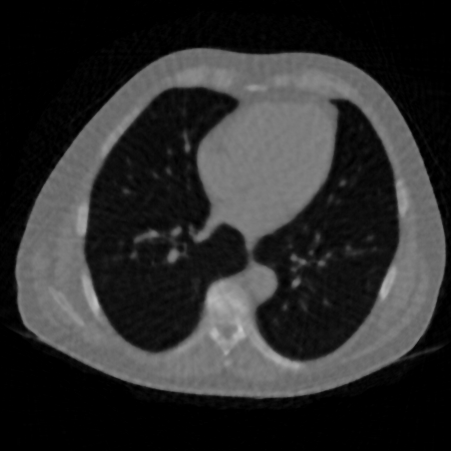}};
					\spy on (0.6,-0.60) in node [left] at (-0.3,-1);   
				\end{scope}
			\end{tikzpicture}
		\end{tabular}	
	}
	\hspace{-0.08\linewidth}
	\subcaptionbox{\Ac{JTV} \label{sugfig:clin_jtv}}
	{
		\begin{tabular}{c}
			\begin{tikzpicture}
				\begin{scope}[spy using outlines={rectangle,yellow,magnification=1.25,size=10mm,connect spies}]
					\node {\includegraphics[viewport=15 50 295 330, clip, width=2.3cm, height=2.3cm]{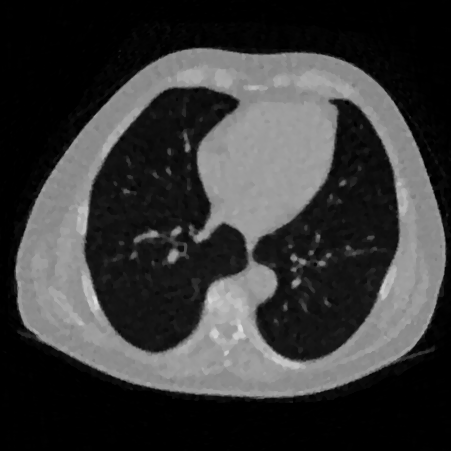}};
					\spy on (0.6,-0.60) in node [left] at (-0.3,-1);  
				\end{scope}
			\end{tikzpicture} \\
			\begin{tikzpicture}
				\begin{scope}[spy using outlines={rectangle,yellow,magnification=1.25,size=10mm,connect spies}]
					\node {\includegraphics[viewport=15 50 295 330, clip, width=2.3cm, height=2.3cm]{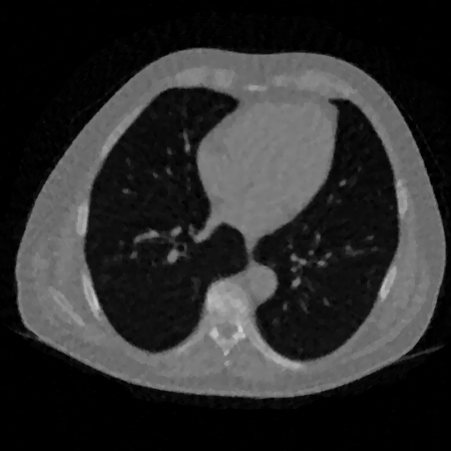}};
					\spy on (0.6,-0.60) in node [left] at (-0.3,-1);   
				\end{scope}
			\end{tikzpicture}
		\end{tabular}	
	}
	
	\caption{
		Comparison of reconstructed clinical data from different reconstruction methods for sparse-view \ac{CT} with top row corresponding to high energy $E_1 = 140$ keV and bottom row to low energy $E_2 = 70$ keV: \subref{sugfig:clin_gt} \ac{GT} image, \subref{sugfig:clin_lbfgs} minimization of the \ac{NLL} function without prior, \subref{sugfig:clin_caol} \ac{CAOL} reconstruction, \subref{sugfig:clin_mcaol} \ac{MCAOL} joint reconstruction, \subref{sugfig:clin_tv} separate reconstruction using \ac{TV} prior and \subref{sugfig:clin_jtv} joint reconstruction using \ac{JTV} prior. 
	}
	\label{fig:CTreconCLIN}
	
\end{figure}

\endgroup

\begin{figure}[!h]
	\centering
	\subcaptionbox{$70$ keV. \label{subfig:Bias_vs_std_clin_e1}}
	{
		\begin{tikzpicture}[scale=.75] 
			\begin{axis}[
				xlabel={$\+{STD}$},
				ylabel={$\+{AbsBias}$},
				grid = major,
				legend columns=2,
				legend cell align=left,
				legend entries={MCAOL,CAOL, TV, JTV},
				legend style={legend pos=north west}
				]
				\addplot[color=blue,mark=*] table[x=xMCAOL, y=yMCAOL] {Fig_9a.txt};
				\addplot[color=black,mark=square*] table[x=xCAOL, y=yCAOL] {Fig_9a.txt};
				\addplot[color=red,mark=diamond*] table[x=xTV, y=yTV] {Fig_9a.txt};
				\addplot[color=green,mark=pentagon*] table[x=xJTV, y=yJTV] {Fig_9a.txt};				
			\end{axis}
		\end{tikzpicture}
	}
	\subcaptionbox{$140$ keV. \label{subfig:Bias_vs_std_clin_e2}}
	{
		\begin{tikzpicture}[scale=.75] 
			\begin{axis}[
				xlabel={$\+{STD}$},
				ylabel={$\+{AbsBias}$},
				grid = major,
				legend columns=2,
				legend cell align=left,
				legend entries={MCAOL, CAOL, TV, JTV},
				]
				\addplot[color=blue,mark=*] table[x=xMCAOL, y=yMCAOL] {Fig_9b.txt};
				\addplot[color=black,mark=square*] table[x=xCAOL, y=yCAOL] {Fig_9b.txt};
				\addplot[color=red,mark=diamond*] table[x=xTV, y=yTV] {Fig_9b.txt};
				\addplot[color=green,mark=pentagon*] table[x=xJTV, y=yJTV] {Fig_9b.txt};	
			\end{axis}
		\end{tikzpicture}
	}
	\caption{Plot of the mean absolute bias ($\+{AbsBias}$) versus the standard deviation ($\+{STD}$) for the  clinical data at \subref{subfig:Bias_vs_std_clin_e1} low X-ray source energy ($70$~keV) and \subref{subfig:Bias_vs_std_clin_e2} high X-ray source energy ($140$ keV).} \label{fig:Bias_vs_std_clin}
	
\end{figure}

In Fig.~\ref{fig:CTreconCLIN} the \ac{GT} image and the reconstruction images for both energies and the different methods are shown; it is worth noting that the \ac{MCAOL} reconstruction (Fig.~\ref{sugfig:clin_mcaol}) is less noisy than the \ac{CAOL} reconstruction (Fig.~\ref{sugfig:clin_caol}).

Figs.~\ref{subfig:Bias_vs_std_clin_e1} and \ref{subfig:Bias_vs_std_clin_e2} report the $\+{AbsBias}$ versus the $\+{STD}$ plots and we obtain a similar behavior compared to the \ac{XCAT} simulations; although the relative distance of the $\+{AbsBias}$ among the simulated algorithms is reduced, \ac{MCAOL} is still outperforming all other methods constantly either by fixing $\+{AbsBias}$ or $\+{STD}$, while the performance of \ac{CAOL} is improving and it is close to the \ac{JTV} solution accuracy.

\subsection{Results for Low-Dose DECT}

We conducted \ac{DECT} reconstruction on the same set of data as in Section~\ref{subsec:res_clin} but with a different \ac{CT} acquisition setup; we substantially decreased the initial photon counts to $\bar{S}_e = 10^3$ (reduction of 2 orders of magnitude compared to the previous experiments) and we doubled the number of views to 120. By approximating the total delivered X-ray dose as the product of the photons intensity times the number of views, it turns out that this scenario is considerably more challenging in terms of ill-posed problem with a total dose reduction of $50$ times.

\begingroup
\tabcolsep = 8pt
\def\arraystretch{0} 	

\begin{figure}[!ht]
	\centering
	\subcaptionbox{No prior \label{sugfig:clin_lbfgs_ld}}
	{
		\begin{tabular}{c}
			\begin{tikzpicture}
				\begin{scope}[spy using outlines={rectangle,yellow,magnification=1.25,size=10mm,connect spies}]
					\node {\includegraphics[viewport=15 50 295 330, clip, width=2.3cm, height=2.3cm]{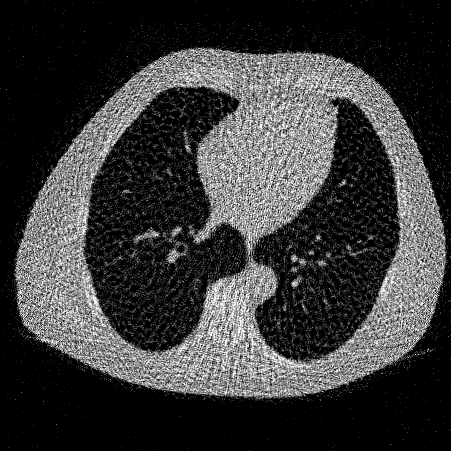}};
					\spy on (0.6,-0.60) in node [left] at (-0.3,-1);   
				\end{scope}
			\end{tikzpicture} \\
			\begin{tikzpicture}
				\begin{scope}[spy using outlines={rectangle,yellow,magnification=1.25,size=10mm,connect spies}]
					\node {\includegraphics[viewport=15 50 295 330, clip, width=2.3cm, height=2.3cm]{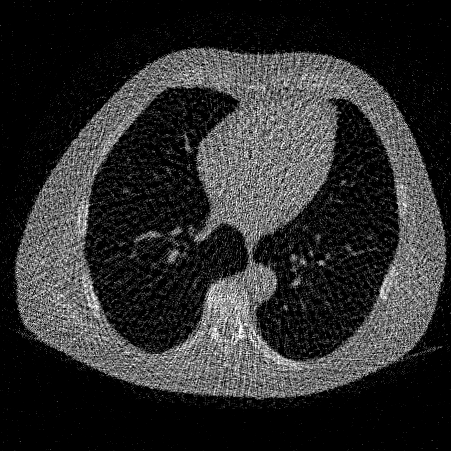}};
					\spy on (0.6,-0.60) in node [left] at (-0.3,-1);   
				\end{scope}
			\end{tikzpicture}
		\end{tabular}
		
	}
	\hspace{-0.08\linewidth}
	\subcaptionbox{\Ac{CAOL} \label{sugfig:clin_caol_ld}}
	{
		\begin{tabular}{c}
			\begin{tikzpicture}
				\begin{scope}[spy using outlines={rectangle,yellow,magnification=1.25,size=10mm,connect spies}]
					\node {\includegraphics[viewport=15 50 295 330, clip, width=2.3cm, height=2.3cm]{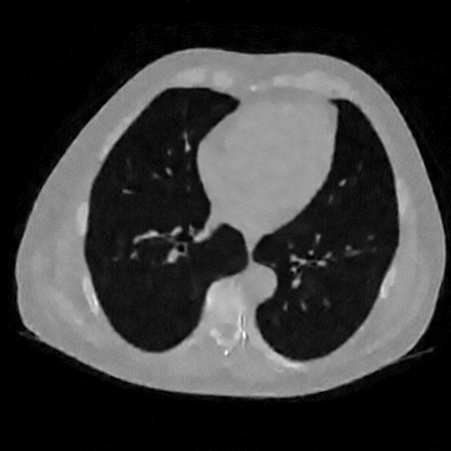}};
					\spy on (0.6,-0.60) in node [left] at (-0.3,-1);  
				\end{scope}
			\end{tikzpicture} \\
			\begin{tikzpicture}
				\begin{scope}[spy using outlines={rectangle,yellow,magnification=1.25,size=10mm,connect spies}]
					\node {\includegraphics[viewport=15 50 295 330, clip, width=2.3cm, height=2.3cm]{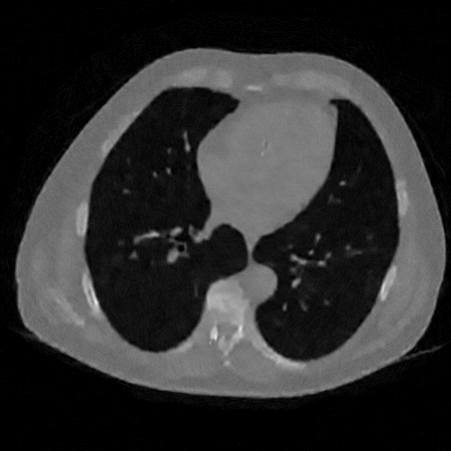}};
					\spy on (0.6,-0.60) in node [left] at (-0.3,-1);  
				\end{scope}
			\end{tikzpicture}
		\end{tabular}	
	}
	\hspace{-0.08\linewidth}
	\subcaptionbox{\Ac{MCAOL} \label{sugfig:clin_mcaol_ld}}
	{
		\begin{tabular}{c}
			\begin{tikzpicture}
				\begin{scope}[spy using outlines={rectangle,yellow,magnification=1.25,size=10mm,connect spies}]
					\node {\includegraphics[viewport=15 50 295 330, clip, width=2.3cm, height=2.3cm]{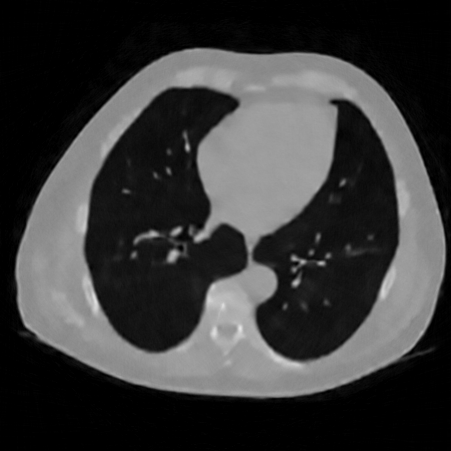}};
					\spy on (0.6,-0.60) in node [left] at (-0.3,-1);  
				\end{scope}
			\end{tikzpicture}\\
			\begin{tikzpicture}
				\begin{scope}[spy using outlines={rectangle,yellow,magnification=1.25,size=10mm,connect spies}]
					\node {\includegraphics[viewport=15 50 295 330, clip, width=2.3cm, height=2.3cm]{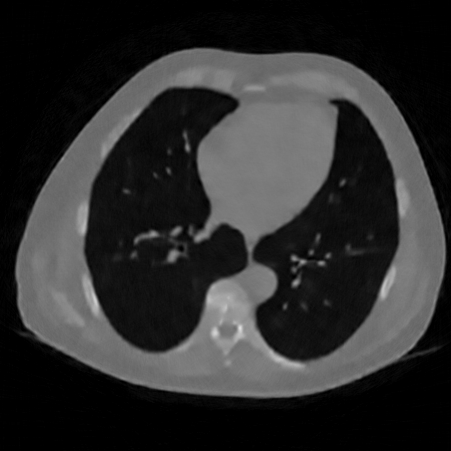}};
					\spy on (0.6,-0.60) in node [left] at (-0.3,-1);  
				\end{scope}
			\end{tikzpicture}
		\end{tabular}	
	}
	\hspace{-0.08\linewidth}
	\subcaptionbox{\Ac{TV} \label{sugfig:clin_tv_ld}}
	{
		\begin{tabular}{c}
			\begin{tikzpicture}
				\begin{scope}[spy using outlines={rectangle,yellow,magnification=1.25,size=10mm,connect spies}]
					\node {\includegraphics[viewport=15 50 295 330, clip, width=2.3cm, height=2.3cm]{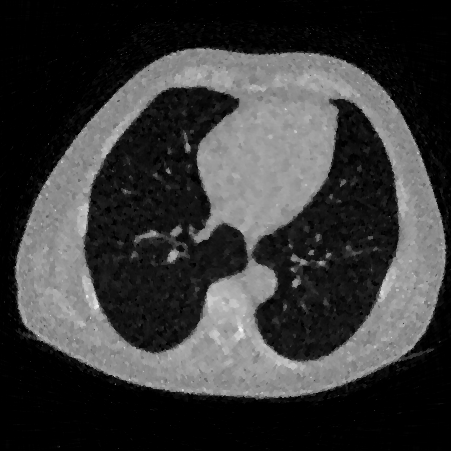}};
					\spy on (0.6,-0.60) in node [left] at (-0.3,-1);  
				\end{scope}
			\end{tikzpicture} \\
			\begin{tikzpicture}
				\begin{scope}[spy using outlines={rectangle,yellow,magnification=1.25,size=10mm,connect spies}]
					\node {\includegraphics[viewport=15 50 295 330, clip, width=2.3cm, height=2.3cm]{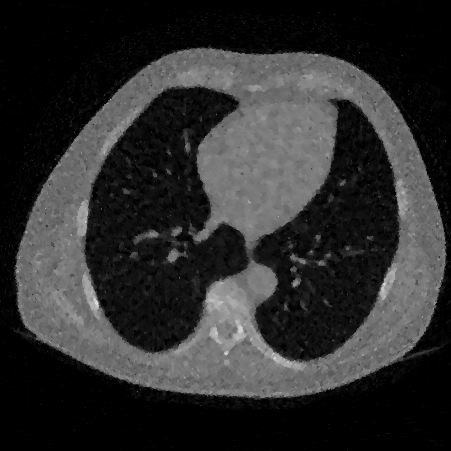}};
					\spy on (0.6,-0.60) in node [left] at (-0.3,-1);  
				\end{scope}
			\end{tikzpicture}
		\end{tabular}	
	}
	\hspace{-0.08\linewidth}
	\subcaptionbox{\Ac{JTV} \label{sugfig:clin_jtv_ld}}
	{
		\begin{tabular}{c}
			\begin{tikzpicture}
				\begin{scope}[spy using outlines={rectangle,yellow,magnification=1.25,size=10mm,connect spies}]
					\node {\includegraphics[viewport=15 50 295 330, clip, width=2.3cm, height=2.3cm]{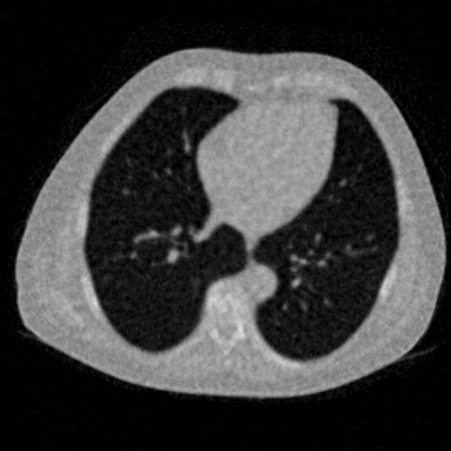}};
					\spy on (0.6,-0.60) in node [left] at (-0.3,-1);   
				\end{scope}
			\end{tikzpicture}\\
			\begin{tikzpicture}
				\begin{scope}[spy using outlines={rectangle,yellow,magnification=1.25,size=10mm,connect spies}]
					\node {\includegraphics[viewport=15 50 295 330, clip, width=2.3cm, height=2.3cm]{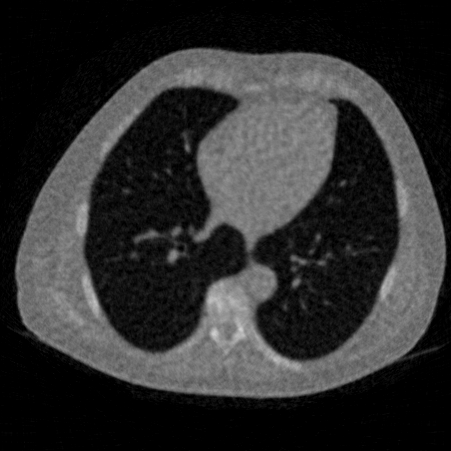}};
					\spy on (0.6,-0.60) in node [left] at (-0.3,-1);  
				\end{scope}
			\end{tikzpicture}
		\end{tabular}	
	}
	\hspace{-0.08\linewidth}
	\subcaptionbox{\Ac{CAOL}-\ac{PWLS} \label{sugfig:clin_caol_pwls_ld}}
	{
		\begin{tabular}{c}
			\begin{tikzpicture}
				\begin{scope}[spy using outlines={rectangle,yellow,magnification=1.25,size=10mm,connect spies}]
					\node {\includegraphics[viewport=15 50 295 330, clip, width=2.3cm, height=2.3cm]{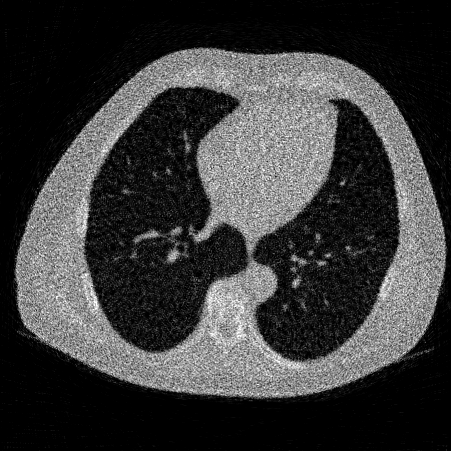}};
					\spy on (0.6,-0.60) in node [left] at (-0.3,-1);  
				\end{scope}
			\end{tikzpicture}\\
			\begin{tikzpicture}
				\begin{scope}[spy using outlines={rectangle,yellow,magnification=1.25,size=10mm,connect spies}]
					\node {\includegraphics[viewport=15 50 295 330, clip, width=2.3cm, height=2.3cm]{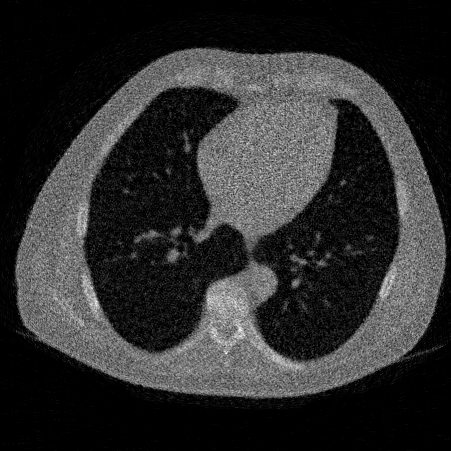}};
					\spy on (0.6,-0.60) in node [left] at (-0.3,-1);   
				\end{scope}
			\end{tikzpicture}
		\end{tabular}	
	}
	
	\caption{
		Comparison of reconstructed clinical data from different reconstruction methods for low-dose \ac{CT} with top row corresponding to high energy $E_1 = 140$ keV and bottom row to low energy $E_2 = 70$ keV: \subref{sugfig:clin_lbfgs_ld} minimization of the \ac{NLL} cost function without prior, \subref{sugfig:clin_caol_ld} \ac{CAOL} reconstruction, \subref{sugfig:clin_mcaol_ld} \ac{MCAOL} joint reconstruction, \subref{sugfig:clin_tv_ld} separate reconstruction using \ac{TV} prior, \subref{sugfig:clin_jtv_ld} \ac{JTV} prior and \subref{sugfig:clin_caol_pwls_ld} \ac{CAOL}-\ac{PWLS} reconstruction.
	}
	\label{fig:CTreconCLIN_lowdose}
	
\end{figure}

\endgroup

\begin{figure}[!h]
	\centering
	\subcaptionbox{$70$ keV. \label{subfig:Bias_vs_std_clin_ld_e1}}
	{
		\begin{tikzpicture}[scale=.75] 
			\begin{axis}[
				xlabel={$\+{STD}$},
				ylabel={$\+{AbsBias}$},
				grid = major,
				legend columns=2,
				legend cell align=left,
				legend entries={MCAOL, CAOL, TV, JTV, CAOL-PWLS},
				legend style={at={(0.8,1.05)},anchor=north}
				]
				\addplot[color=blue,mark=*] table[x=xMCAOL, y=yMCAOL] {Fig_11a.txt};
				\addplot[color=black,mark=square*] table[x=xCAOL, y=yCAOL] {Fig_11a.txt};
				\addplot[color=red,mark=triangle*] table[x=xTV, y=yTV] {Fig_11a.txt};
				\addplot[color=green,mark=square*] table[x=xJTV, y=yJTV] {Fig_11a.txt};	
				\addplot[color=magenta,mark=pentagon*] table[x=xCAOLPWLS, y=yCAOLPWLS] {Fig_11a.txt};	       
			\end{axis}
		\end{tikzpicture}
	}
	\subcaptionbox{$140$ keV. \label{subfig:Bias_vs_std_clin_ld_e2}}
	{
		\begin{tikzpicture}[scale=.75] 
			\begin{axis}[
				xlabel={$\+{STD}$},
				ylabel={$\+{AbsBias}$},
				grid = major,
				legend columns=2,
				legend cell align=left,
				legend entries={MCAOL, CAOL, TV, JTV, CAOL-PWLS},
				legend style={at={(0.8,1.05)},anchor=north} 
				]
				\addplot[color=blue,mark=*] table[x=xMCAOL, y=yMCAOL] {Fig_11b.txt};
				\addplot[color=black,mark=square*] table[x=xCAOL, y=yCAOL] {Fig_11b.txt};
				\addplot[color=red,mark=triangle*] table[x=xTV, y=yTV] {Fig_11b.txt};
				\addplot[color=green,mark=square*] table[x=xJTV, y=yJTV] {Fig_11b.txt};	
				\addplot[color=magenta,mark=pentagon*] table[x=xCAOLPWLS, y=yCAOLPWLS] {Fig_11b.txt};	
			\end{axis}
		\end{tikzpicture}
	}
	\caption{Plot of the mean absolute bias ($\+{AbsBias}$) versus the standard deviation ($\+{STD}$) for the low-dose ($I_0 = 10^3$) reconstruction with clinical data at \subref{subfig:Bias_vs_std_clin_ld_e1} low X-ray source energy ($70$~keV) and \subref{subfig:Bias_vs_std_clin_ld_e2} high X-ray source energy ($140$ keV).} \label{fig:Bias_vs_std_clin_ld}
	
\end{figure}

We used this simulation to prove that \ac{MCAOL} returns a more accurate solution compared to other priors. Furthermore, we prove that despite the higher computational complexity to minimize the exact Poisson \ac{NLL} in \eref{eq:NLL_MCAOLPoisson} compared to solving the problem with a weighted least-squares approximated \ac{NLL}, i.e., \ac{PWLS} data-fit cost function, \ac{MCAOL} achieves substantial improved bias accuracy compared to the \ac{PWLS} solution. 
To perform these experiments, we used the same optimal learned convolutional filters as obtained by the \ac{MCAOL} training procedure detailed in Section~\ref{subsec:res_clin} and the \ac{GT} images in Fig.~\ref{fig:CTreconCLIN}(a). 

Fig.~\ref{fig:CTreconCLIN_lowdose} show the reconstruction images for both energies and different methods; \ac{MCAOL} (Fig.~\ref{sugfig:clin_mcaol_ld}) accurately reconstruct the image features compared to all other methods and it is confirmed that the PWLS model performs poorly (Fig.~\ref{sugfig:clin_caol_pwls_ld}).   

Figs.~\ref{subfig:Bias_vs_std_clin_ld_e1} and \ref{subfig:Bias_vs_std_clin_ld_e2} show either that \ac{MCAOL} is consistently outperforming the other methods in terms of accuracy and variance and that the Poisson \ac{NLL} formulation leads to a noticeable improvement compared to the \ac{PWLS} formulation as it is indicated by comparing \ac{CAOL} and \ac{CAOL}-\ac{PWLS}.

%% file: conclusion.tex
\section{Discussion} \label{sec:discussion}

The bias-variance trade-off analysis of the estimation results over the regularization parameters confirms that \ac{MCAOL} allows to achieve the minimum absolute bias compared to \ac{CAOL} and other \ac{MBIR} state-of-the-art methods and also reduce standard deviation. 
Furthermore, \ac{MCAOL} has the benefit of requiring less memory respect to \ac{DL} methods because of the convolutional structure of the trained filters.

The \ac{MCAOL} algorithm can be potentially extended to multi-energies $e=1,\ldots, E$ since the $l_{1,0}$ semi-norm in Eq.~(\ref{eq:l_1_0}) can be defined for a set of $E$-vectorised feature maps $\*z_1, \ldots, \*z_E$. Each $\*z_e, e=1, \ldots,E$ is a column vector of dimension $J\times 1$. Then the joint $l_{1,0}$ semi-norm is defined as
\begin{equation}
	\| (\*z_1, \ldots, \*z_E) \|_{1,0} = \sum_{j=1}^J\*1_{]0, +\infty[}\left(|z_{1,j}| + \ldots + |z_{E,j}|\right) 
\end{equation}

\noindent While the statistical noise tends to be higher in the multi-energy case, on each sub-band the contribute of the noise is reduced since the noise is split on more energy bands. Therefore, we believe that evaluating the joint norm, i.e., non-zeros elements in the feature vectors in overlapping positions for all energies at the same time, will reduce the degradation due to the increased overall noise.

The \ac{MCAOL} framework allows to utilize any mixed norms for the jointly sparse regularization and other norms, such as the $l_{2,1}$-norm which as proposed by \citeasnoun{degraux2017online} for convolutional synthesis operator learning, may also be considered.  

In our experiments we have considered the product between the X-ray source intensity and the number of projection angles as an empirical measure for the total transmitted X-ray dose. While this metric gives a good approximation of the dose, we consider the analysis of the standardized measure of radiation dose, i.e., CT dose index (CTDI), as well as the absorbed dose as a follow-up study.

We account the open problems of how to optimally select both the regularization norm and regularization parameter according to the dataset for future algorithm development. 

Although this work focuses on the multi-channel imaging reconstruction problem, we believe that our proposed method can be utilized in conjunction to \ac{DECT} to task-oriented material decomposition problems. In particular, while an approach would be to design a material decomposition module in the image space which takes as input the \ac{MCAOL} reconstructed images, a more compelling strategy would be designing a direct approach from sinograms to material images through \ac{MCAOL}. 

Furthermore, \ac{MCAOL} method can be exploited for other multimodal imaging application such as PET/CT and PET/MRI. In the multimodal case, given the different intensity range on each channel, a further analysis on how to choose the \ac{NLL} weights $\gamma_1\neq\gamma_2$ in (\ref{eq:MPML}) should be conducted to properly balancing the information coming from the different modalities. 

Finally, from a learning point of view, \ac{MCAOL} training can be seen as a multi-channel single layer unsupervised convolutional autoencoder \cite[Appendix A]{Chun2019} which paves the way to extend this approach to deeper autoencoder architectures to capture more complex features such as textures. 
The analysis and comparison of the proposed \ac{MCAOL} approach with other supervised deep learning approaches is planned as a follow-up study. It is important to stress that \ac{MCAOL} inherits a precise mathematical derivation and therefore it should not be susceptible of instabilities in the reconstruction which have been proven to occur with deep learning methods \cite{antun2020instabilities}.

We consider these problems as future development of the proposed algorithm.

\section{Conclusion}\label{sec:conclusion}

In this work, we have extended the convolutional analysis operator framework to multi-channel imaging and we have applied and extensively analyzed the proposed method to the \ac{DECT} application. The presented results show that by using the information coming from both energies and allowing the channels to ``talk to each other'' a more accurate solution of the reconstruction problem can be achieved together with a reduction of the noise in the estimate. The coupling between energies is encapsulated by using an $l_{1,0}$ sparse mixed norm in the \ac{MCAOL} optimization problems both for training and reconstruction. We obtain consistently better performances across different \ac{DECT} acquisition scenarios from sparse-views to low-dose photon counts.